\documentclass{aa}  

\usepackage{graphicx}
\usepackage{txfonts}
\usepackage{lipsum}
\usepackage{subcaption}
\usepackage{lscape}
\usepackage{placeins}
\usepackage{csquotes}
\usepackage{hyperref}
\usepackage{tabularx}
\usepackage{adjustbox}
\usepackage{float}
\usepackage{geometry}
\usepackage[normalem]{ulem}
\usepackage{multicol}

\begin{document}

   \title{Globular clusters in \textsc{OrbIT}: complete dynamical characterisation of the globular cluster population of the Milky Way through updated orbital reconstruction}

   \author{M. De Leo\inst{1,2,3,4}\thanks{micheledl89@gmail.com}
        \and M. Zoccali\inst{3,4}
        \and J. Olivares-Carvajal\inst{3,4}
        \and B. Acosta-Tripailao\inst{3,4}
        \and F. Gran\inst{5}
        \and R. Contreras-Ramos\inst{3,4}
        }

   \institute{Dipartimento di Fisica e Astronomia, Universit\`{a} degli Studi di Bologna, Via Piero Gobetti 93/2, Bologna, 40129, Italy
            \and Osservatorio di Astrofisica e Scienza dello Spazio di Bologna, INAF, Via Piero Gobetti 93/3, Bologna, 40129, Italy
            \and Instituto de Astrof\'{i}sica, Pontificia Universidad Cat\'{o}lica de Chile, Av. Vicu\~{n}a Mackenna 4860, 782-0436, Macul, Santiago, Chile           
            \and Instituto Milenio de Astrof\'{i}sica MAS, Av. Vicu\~{n}a Mackenna 4860, 782-0436, Macul, Santiago, Chile           
            \and Universit\'{e} C\^{o}te d’Azur, Observatoire de la C\^{o}te d’Azur, CNRS, Laboratoire Lagrange, 06304 Nice, France}

   \date{Received xxx}
 
  \abstract
   {In hierarchical structure formation, the content of a galaxy is determined both by its in-situ processes and by material added via accretions. Globular clusters in particular represent a window for the study of the different merger events that a galaxy underwent. Establishing the correct classification of in-situ and accreted tracers, and distinguishing the various different progenitors that contributed to the accreted population are important tools to deepen our understanding of galactic formation and evolution.}
   {Our aim is to refine our knowledge of the assembly history of the Milky Way by studying the dynamics of its globular cluster population and establishing an updated classification among in-situ objects and the different merger events identified.} 
   {We used a custom built orbit integrator to derive precise orbital parameters, integrals of motions and adiabatic invariants for the globular cluster sample studied. By properly accounting for the rotating bar, which transforms the underlying model in a time-varying potential, we proceeded to a complete dynamical characterisation of the globular clusters.}
   {We present a new catalogue of clear associations between globular clusters and structures (both in-situ and accreted) in the Milky Way, and a full table of derived parameters. By using all dynamical information available, we were able to attribute previously unassociated or misclassified globular clusters to the different progenitors, including those responsible for the Aleph, Antaeus, Cetus, Elqui, and Typhon merger events.}
   {By using a custom built orbit integrator and properly accounting for the time-varying nature of the Milky Way potential, we have shown the depth of information that can be extracted from a purely dynamical analysis of the globular clusters of our Galaxy. By merging our dynamical analysis with complementary chronochemical studies, we will be able to uncover the remaining secrets of the accretion history of the Milky Way.}


    \titlerunning{Globular clusters in \textsc{OrbIT}: dynamical characterisation of MW GCs}
   \maketitle

\section{Introduction}\label{intro}
In the prevailing paradigm of $\Lambda$ Cold Dark Matter ($\Lambda$CDM) the structure of the Universe is believed to have formed hierarchically \citep{1965ApJ...142.1317P, 1970ApJ...162..815P}, with the largest overdensities accreting smaller neighbours and becoming the present-day massive galaxies \citep{1978MNRAS.183..341W, 2006MNRAS.371..885R}. This turbulent process leaves traces that can be studied and decoded in the stellar and globular cluster (GC) populations of the final surviving galaxy and these traces allow to identify the properties and contribution of the populations formed \textquote{in-situ} and of the accreted material from the mergers \citep{1962ApJ...136..748E, 1978ApJ...225..357S, 2005ApJ...635..931B}. Our Galaxy, the Milky Way (MW), provides a privileged window through which we can peer at the processes shaping galactic evolution and formation by studying the stellar and GC populations present. Due to the complexities of stellar and cluster formation, GCs offer a better opportunity of studying the merger history and accreted content of a galaxy \citep[e.g. ][ and references therein]{2024MNRAS.528.3198B}. To this end, it is of paramount importance to discriminate between in-situ and accreted GCs. First studies aimed at distinguishing the two families used available chemical and spatial information \citep{1985ApJ...293..424Z, 1996ASPC...92..211Z}, later the availability of more accurate chemical data and age estimations opened up the way for the use of Age-Metallicity Relations \citep[AMR, ][]{2009ApJ...694.1498M, 2010MNRAS.404.1203F, 2013MNRAS.436..122L}. Running in parallel to the chronochemical techniques, the availability of samples with complete position and velocity information allowed the use of Integrals of Motions (IoM) and adiabatic invariants (such as the actions), conserved after the mergers, to isolate substructures in dynamical parameter spaces \citep{1999MNRAS.307..495H, 2000MNRAS.319..657H, 2008gady.book.....B, 2010MNRAS.408..935G}. Concurrently, the \textquote{accreted} family was further split up with more detailed identifications of the first confirmed mergers, Sagittarius \citep{1994Natur.370..194I} and the so-called \textquote{Helmi Streams} \citep{1999Natur.402...53H}. The floodgates of discovery were opened with the data releases of the \textit{Gaia} mission \citep{2016A&A...595A...1G}, providing position and velocity information for millions of targets, and with complementary wide spectroscopic surveys \citep[chief among them the Apache Point Observatory Galactic Evolution Experiment, APOGEE, ][]{2017AJ....154...94M}. These led to the identification of several different candidate merger events, a research that is still ongoing due to the continuous discovery of new structures. The seminal work of \citet{2019A&A...630L...4M} set off systematic efforts to classify the merger events, characterise them in the different parameter spaces, and establish affiliations between the MW GCs and the different progenitors. Refined research techniques using clustering algorithms \citep{2022A&A...665A..57L, 2022A&A...665A..58R, 2023A&A...670L...2D} enabled the identification of many new substructures. Robust spectroscopic and photometric analysis allow ever more precise estimations of the age and chemical composition of stars and GCs \citep{2023MNRAS.520.5671H, 2023A&A...680A..20M, 2024A&A...684A..37C} facilitating the establishment of links with the different progenitors. One often overlooked piece of the puzzle is the potential model used to derive the orbital parameters, the IoMs and the adiabatic invariants. To properly capture the inherent intricacies of the structure of the MW it is necessary to incorporate all observed components of the Galaxy. In particular, the rotating bar is a crucial player both for the direct effect it has on the inner regions of the MW and for the time-varying nature of its action that creates resonances \citep{2021MNRAS.500.4710C, 2021MNRAS.505.2412C, 2022MNRAS.509..844T, 2024MNRAS.532.4389D}. We have developed the \textsc{Orbital Integration Tool (OrbIT)} as a flexible, transparent and efficient code to perform orbit integrations with static and time-varying MW potentials. \textsc{OrbIT} has been employed in \citet{2024A&A...687A.312O} to study the RR Lyrae bulge population and this paper presents our first results on the MW GCs. This work is but a stepping stone towards a deeper understanding of how perturbations and time-varying components of the MW potentials shape and affect the inventory and structure of our Galaxy.

In Section ~\ref{data} we present the data catalogues used, Section ~\ref{method} contains an in-depth description of \textsc{OrbIT} and the underlying potential model it uses, Section ~\ref{results} contains our results, which we discuss in Section ~\ref{discu} while Section ~\ref{conclusion} has a summary of our conclusions.

\section{Data}\label{data}

The main source of the data used throughout this work is the extensive repository of GC parameters by Baumgardt et al\footnote{\href{https://people.smp.uq.edu.au/HolgerBaumgardt/globular/}{https://people.smp.uq.edu.au/HolgerBaumgardt/globular/}} \citep{2017MNRAS.464.2174B, 2018MNRAS.478.1520B, 2021MNRAS.505.5957B, 2021MNRAS.505.5978V}. For all the GCs in the Gran \textquote{family} we use the MUSE \citep{2006NewAR..49..618S, 2010SPIE.7735E..08B, 2012SPIE.8447E..37S, 2020SPIE11448E..0VH} radial velocities reported by \citet{2024A&A...683A.167G} in place of those reported in the Baumgardt repository. We do the same with VVV-CL001, using the radial velocity from MUSE data reported in \citet{2022MNRAS.513.3993O}. Likewise, for Patchick 126 and VVV-CL160, we use the proper motions and radial velocities measured with a combination of Gaia DR3 \citep{2023A&A...674A...1G}, VVV-X \citep{2010NewA...15..433M, 2012A&A...537A.107S} and IGRINS \citep{2014SPIE.9147E..1DP, 2018SPIE10702E..0QM} by \citet{2023A&A...669A.136G}. Finally, we complement the sample with a few additions from more recent works: VVV-CL002 \citep{2011A&A...535A..33M, 2021A&A...648A..86M, 2024A&A...683A.150M}, 
Gran 4 \citep{2022MNRAS.509.4962G, 2024A&A...683A.167G}, ESO 93-08 \citep{2023A&A...669A.136G}, and Patchick 122 \citep{2022A&A...659A.155G, 2023A&A...669A.136G}.

We transform the observed positions, distances and 3D velocities to the Galactocentric reference frame assuming the velocity vector of the Sun to be $(U_{\odot}\,,V_{\odot}\,,W_{\odot})=(11.1\,,12.24\,,7.25) \ {\rm km \ s^{-1}}$ \citep{2010MNRAS.403.1829S}, the velocity of the Local Standard of Rest $V_{LSR}=220 \ {\rm km \ s^{-1}}$ \citep{2008IAUS..248..141R}, the distance of the Sun from the Galactic Center $R_{\odot}=8.20 \ {\rm kpc}$ \citep{2019A&A...625L..10G}, and the height of the Sun above the Galactic Plane $z_{\odot}=20.8 \ \rm{ pc}$ \citep{2019MNRAS.482.1417B}.

The vast majority ($90\%$) of the sample has observational errors on distances, radial velocities and proper motions along the $\alpha$ and $\delta$ directions of, respectively, less than $1 {\rm \ kpc}, \ 2 {\rm \ km/s}, \ 0.1 {\rm \ mas/yr}, \ {\rm and} \ 0.05 {\rm \ mas/yr}$. We tested that the observational errors have a negligible impact on the classification of the GCs inside the various progenitor families and show in Appendix~\ref{obserrs} that this is true even for the GCs with the highest uncertainties on distance (AM 1, Eridanus, Palomar 3 and Palomar 4) and kinematical values (2MASS-GC01, AM 1 and Patchick 122).

\section{Method}\label{method}

The study of the GC population in this paper is conducted mainly using the custom-made code \textsc{OrbIT}. As the name implies, \textsc{OrbIT} is a code for orbit integrations and was designed to be customizable based on the needs of the problem to be tackled. The decision to develop and use a custom code in place of using a well established one like \textsc{galpy} \citep{2015ApJS..216...29B} or \textsc{AGAMA} \citep{2019MNRAS.482.1525V} stems from the desire to have complete control over the Galactic model used, on every step of the orbit integration and on the available outputs. This allows us to gauge sources of errors as transparently and completely as possible and properly evaluate realistic uncertainties on the parameters derived through the integration.

The orbital history of the GCs is computed with \textsc{OrbIT} backwards in time for 10 Gyr (to allow all GCs to complete more than one orbit) with a timestep of $10^4 \ $ yrs. The recovered orbital parameters (see subsection ~\ref{out}) are then used to identify the most likely progenitors for each GC. Starting from the classification of \citet{2019A&A...630L...4M}, we reassess the various orbital families and established memberships, focusing on relatively new GCs and on previously known ones with uncertain ancestry.

\subsection{General information on \textsc{OrbIT}}\label{orbgen}

\textsc{OrbIT} is a code written in the C computing language and it uses a \textquote{leap-frog} integration, specifically of the \textquote{kick-drift-kick} type. This choice is driven by the fact that this is a type of modified Verlet integration \citep{1967PhRv..159...98V,2008gady.book.....B} that fulfills two very important prerequisites for dynamical modeling: it is inherently time-reversible, thus allowing for both backward and forward integrations, and it suppresses numerical errors at each step \citep{1992PhR...216...63R,2003AcNum..12..399H}. Taken together, these characteristics ensure that the IoMs do not drift over time due to errors in the numerical approximation at each step and thus the evolution stays accurate over extremely long integrations made of a large number of steps.

The \textquote{leap-frog kick-drift-kick} scheme first computes the velocities of the tracers at the half-step $n+1/2$, then the positions and accelerations at the full step $n+1$ and then updates the velocities at the full step. This scheme ensures a good approximation of the evolution of the system while only doing one time per step the computationally expensive acceleration calculations. We note here that, while this is optimal for our use case and current model, it might not be the best solution when taking into account effects that explicitly depend on the velocity of tracers, such as dynamical friction for satellite galaxies \citep{1943ApJ....97..255C,1943ApJ....97..263C,1943ApJ....98...54C}. This is because adding dynamical friction (which depends on velocity, so the needed calculations would be performed twice for each timestep) would slow down the integration time. From a dynamical perspective, it has been demonstrated that dynamical friction has a negligible effect on the orbits of most GCs \citep{2022MNRAS.510.5945M} with very few exceptions for the lowest energy ones (which would be classified as bulge GCs anyway).

\textsc{OrbIT} requires as initial conditions the 6D Galactocentric phase space coordinates of the tracers $(X, \ Y , \ Z , \ V_x, \ V_y, \ V_z)_{gal}$. The number of timesteps and number of tracers can be specified in a parameter file without recompiling the source code, the step size can be modified inside the source code and requires recompiling. The source code is designed to compile without errors or warnings on the most common versions of the GNU Compiler Collection (GCC)\footnote{\href{https://gcc.gnu.org/}{https://gcc.gnu.org/}} and has been tested and verified on Windows and Linux operating systems.
The classic sanity check of the conservation of IoMs during integration is reported in Appendix~\ref{appa}

\subsection{Galactic potential model}\label{MWpot}

The model of the MW gravitational potential employed in \textsc{OrbIT} is composed of a Navarro-Frenk-White (NFW) Dark Matter (DM) halo \citep{1996ApJ...462..563N}, two Miyamoto-Nagai (MN) stellar discs \citep{1975PASJ...27..533M}, two MN gaseous discs, a Long-Murali (LM) rotating bar \citep{1992ApJ...397...44L} and a spherical bulge component modeled as an Plummer profile \citep{1911MNRAS..71..460P}. Finally, but with a negligible impact on the final results, the central black hole of our galaxy is present as a point source with mass $M_{BH} = 4.154 \times 10^6 M_{\odot}$ \citep{2019A&A...625L..10G}.

The NFW DM halo is the most massive component of the potential with mass $M_{DM} = 8.0 \times 10^{11} M_{\odot}$, scale radius $r_s = 16$ kpc and halo concentration parameter $c = 15.3$. All the parameter values chosen are quite commonly used, i.e. in the \texttt{MWPotential2014} from \citet{2015ApJS..216...29B}.

The two MN stellar discs are set up to reproduce the observed thin and thick disc of the MW, their scale heights are fixed at $z_{thin} = 0.3$ kpc and $z_{thick} = 0.9$ kpc \citep[i.e.][]{2017MNRAS.465...76M} and their masses and scale lengths are chosen to recover the local mass densities measured by \citet{2022MNRAS.513.4130L}: $M_{thin} = 3.65 \times 10^{10} M_{\odot}$, $M_{thick} = 1.55 \times 10^{10} M_{\odot}$, $a_{thin} = 3.5$ kpc and $a_{thick} = 2.0$ kpc.

The two MN gaseous discs represent the HI and molecular gas discs observed in the MW with their masses (for a total gas mass of $M_{gas} = 1.22 \times 10^{10} M_{\odot}$) and scale heights fixed to the same values as the gas discs in \citet{2017MNRAS.465...76M}, $z_{ \rm HI} = 0.085 {\rm \ kpc \ and \ } z_{ \rm HII} = 0.045 {\rm \ kpc}$. Like their stellar counterparts, their scale lengths are set to reproduce the local mass densities measured for the respective gases (i.e. \citealt{2015ApJ...814...13M}), $a_{ \rm HI} = 1.824 {\rm \ kpc \ and \ } a_{ \rm HII} = 5.895 {\rm \ kpc}$.

The bulge of our model is composed of both a bar and a spherical component. The LM bar has a mass of $M_{bar} = 1.0 \times 10^{10} M_{\odot}$, is geometrically similar to the single-bar model in \citet{2015MNRAS.450.4050W}, with semiaxes $a_{bar} = 5.5\, , \ b_{bar} = 0.68\, , \ c_{bar} = 0.09 {\rm \ kpc}$, an inclination angle $\alpha_{bar} = 30^{\circ}$ and rotating with a pattern speed $\Omega_p = 41.3\pm3$\,km\,s$^{-1}$\,kpc$^{-1}$ \citep{2019MNRAS.488.4552S}. Recent studies \citep[i.e.][]{2024A&A...690A.147H} put the bar formation at 8-10 Gyr ago, we tested that the results of this work don't change between running the integration for 8 Gyr or for 10 Gyr. The exceptions are a few high apocentre GCs (AM 1, Crater, Palomar 3, Palomar 4 and Sagittarius II) that complete only a few orbits in the total integration time, thus the orbital parameters are slightly more accurate with the longer integration. The evolution of the bar during its lifetime, including the change in pattern speed \citep{2021MNRAS.500.4710C} will be the topic of future work.
The spherical component has a mass of $M_{sph} = 1.0 \times 10^{10} M_{\odot}$ and a scale radius of $a_{sph} = 0.3$ kpc. The geometrical parameters of both the bar and the spherical component are fitted to the \citet{2015MNRAS.450.4050W} single-bar model which reproduced the observed density distribution of red clump giants in a boxy/peanut bulge shape.

\begin{figure}
\centering
\includegraphics[width=\hsize]{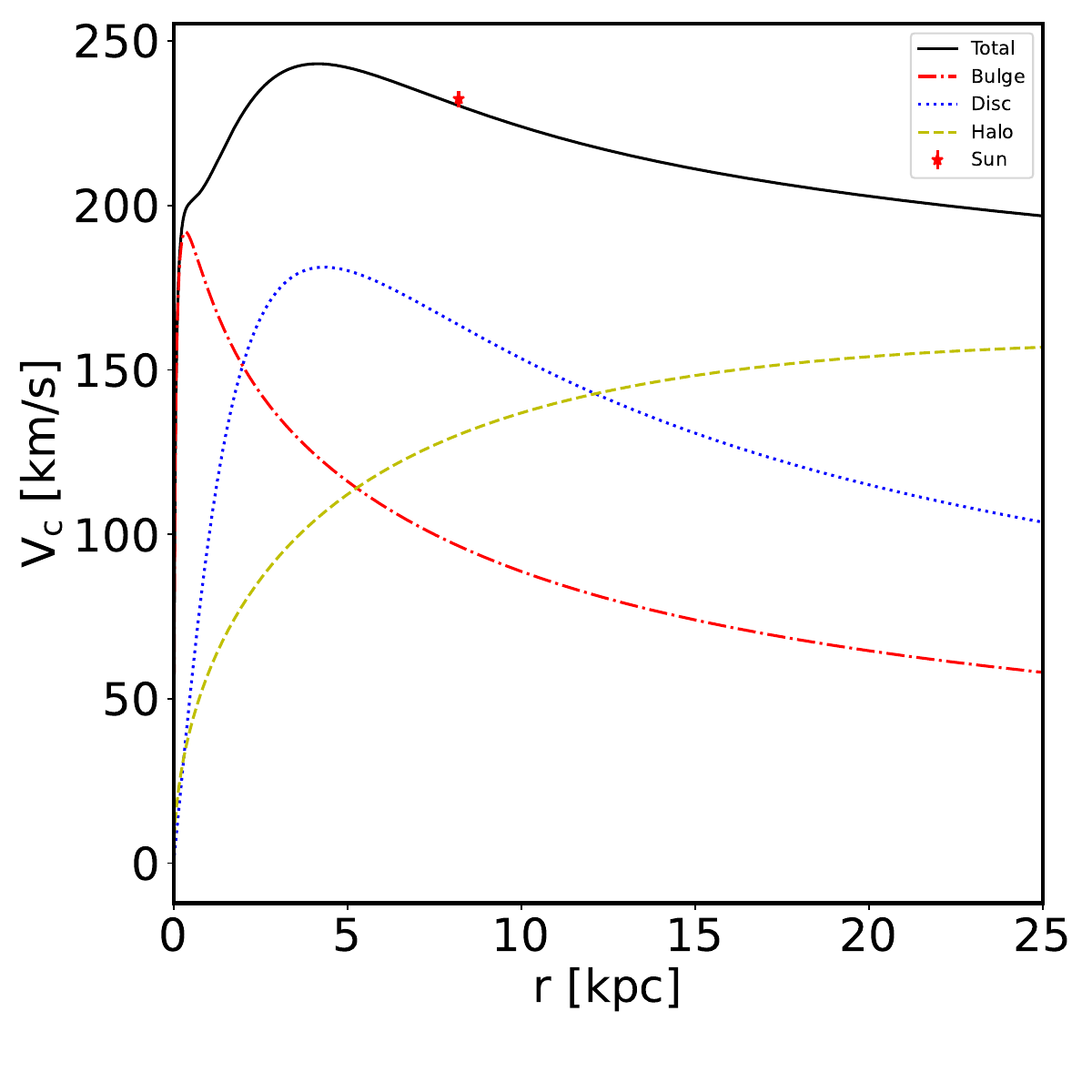}
\caption{Rotational curve of the MW. In solid black the velocity of the full potential model, the yellow dashed line is the contribution from the halo component, the blue dotted line is the total contribution from the discs and the red dash-dotted is the contribution from the bulge (bar and spheroid), the red symbol with errorbar represents the Sun.}
    \label{fig:rot_curve}
\end{figure}

The choice to include a spherical component to the bulge is motivated by the  observations of the MW bulge stellar populations \citep{2016A&A...587L...6V, 2018A&A...618A.147Z, 2021A&A...647A..34L, 2021A&A...656A.156Q, 2024A&A...689A.240Z, 2025A&A...695A.211P}. The total mass of the MW bulge chosen reflects the considerations made in \citet{2020RAA....20..159S} and the equal split between the bar and spherical component follows \citet{2018A&A...618A.147Z}.
Fig.~\ref{fig:rot_curve} shows the rotation curve of the Galaxy produced by our model (including the individual contributions of the different components) which is consistent with the rotation of the Sun \citep[red star with errorbar, from][]{2010MNRAS.403.1829S}.

\subsection{Outputs}\label{out}

\textsc{OrbIT} can be customised to have a wide variety of outputs. In its most common form it provides all the classical orbital parameters, the actions and the circularity of each tracer and it can also output a full orbital history recording the 6D spatial information of each tracer at every timestep.

Given that the underlying potential is non-static, we expect the apocentre, $R_{apo}$, and pericentre, $R_{peri}$, (and therefore the eccentricity, $ecc$) of the tracers to vary in time, making the significance of the recovery of a single value for these parameters quite uncertain in its physical meaning. To work around this fact, the code computes $R_{apo}, \ R_{peri}, \ ecc$ for each complete orbit of each tracer, thus producing a distribution of values. The mean and the standard deviation of each distribution are then, respectively, the final output and the uncertainty for each parameter. Quasi-periodic and more \textquote{well-behaved} tracers will have experienced a smaller variation in the parameter values and have small uncertainties while tracers on chaotic orbits will have larger errors.

The actions $(J_R,J_{\theta},J_{\phi})$ are computed following \citet{2008gady.book.....B}:
\begin{equation}\label{Jphi}
    J_{\phi} = L_{z} \ ,
\end{equation}
\begin{equation}\label{Jtheta}
    J_{\theta} = L - \lvert L_z \rvert \ ,
\end{equation}
\begin{equation}\label{Jr}
    J_R = \frac{2}{\pi}\int_{R_{peri}}^{R_{apo}} \sqrt{2E-2\Phi(r)-\frac{L^2}{r^2}}\rm{d}r \ ,
\end{equation}
where $L$ is the total angular momentum and the integral for $J_R$ is solved numerically with a composite Simpson's 1/3 rule \citep{1991numan.book.....A}.

We exploit the calculation of the actions to compute:
\begin{equation}
    J_{tot} = J_R + J_{\theta} + \lvert J_{\phi} \rvert \ ,
\end{equation}
\begin{equation}
    J_{\perp} = \frac{J_{\theta}-J_R}{J_{tot}} \ ,
\end{equation}
\begin{equation}
    J_{\parallel} = \frac{J_{\phi}}{J_{tot}} \ ,
\end{equation}
these ancillary parameters allow us to explore the \textquote{projected action space map} where $J_{\parallel}$ is the normalised $z$-component of the angular momentum and $J_{\perp}$ gauges the relative importance of radial and vertical motions \citep{2019MNRAS.484.2832V, 2020ApJ...901...48N, 2022ApJ...926..107M}.

The circularity $\varepsilon$ is defined as the ratio of the tracer angular momentum over the angular momentum of a maximally rotating planar orbit having the same specific energy as the tracer \citep{2003ApJ...597...21A, 2014MNRAS.444..515R, 2023MNRAS.525..683O}:
\begin{equation}
    \varepsilon = \frac{L_z}{L_z^{max}(E)} \ ,
\end{equation}
where $L_z^{max}(E)$ is found by identifying the $(V_c\,, \ R_c)$ pair that minimises the following equation for the energy of the tracer:
\begin{equation}
    \bigg\lvert \ \frac{V_c^2}{2} + \Phi(R_c,0) - E \ \bigg\rvert \ ,
\end{equation}
with $V_c$ being the circular velocity defined as:
\begin{equation}
    V_c = \sqrt{R_c \cdot \frac{\delta \Phi}{\delta R}\bigg\rvert_{R_c}} \ ,
\end{equation}

The full list of parameters computed with \textsc{OrbIT} and used throughout this work is as follows: $R_{peri}$, $R_{apo}$, $ecc$, $\lvert Z_{max} \rvert$ (the maximum excursion from the Galactic plane),  $E_{tot}$, $L_x$, $L_y$, $L_z$, $L$, $L_{\perp}$ (the axial, total, and perpendicular angular momenta), $J_{\phi}$, $J_{\theta}$, $J_R$, $J_{tot}$, $J_{\perp}$, $J_{\parallel}$, $\varepsilon$.
Due to the choice of the orientation of the Galactocentric reference frame axes employed in this work, prograde tracers have negative $L_z, \ J_{\phi} \ {\rm and} \ \varepsilon$ while retrograde tracers have positive values of these parameters.

\begin{figure}
\centering
\includegraphics[width=\hsize]{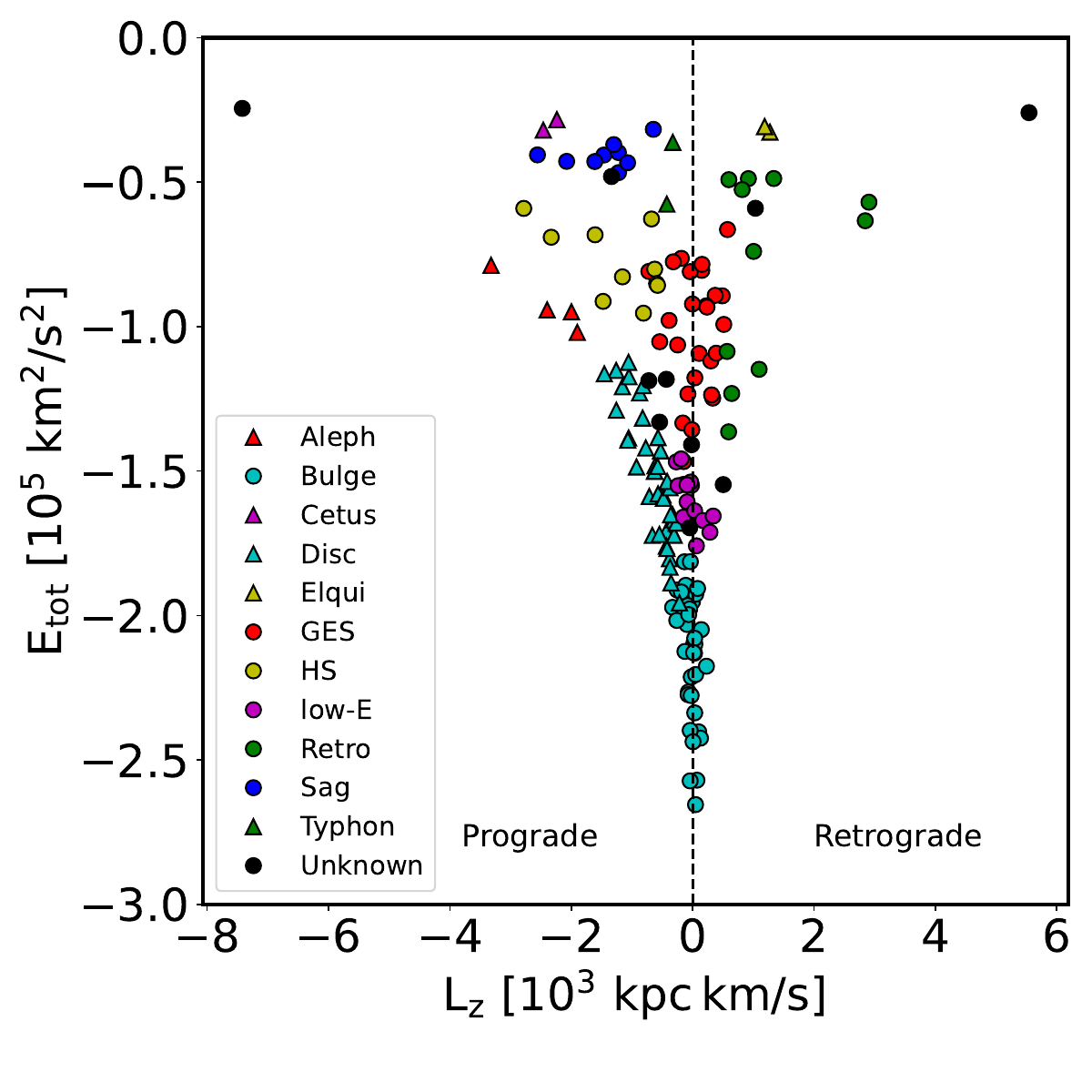}
\caption{The GC sample in the angular momentum versus total energy plane. As per the legend, the GCs are marked based on their affiliation to a progenitor: red triangles for Aleph, cyan circles for the bulge, purple triangles for Cetus, cyan triangles for the disc, yellow triangles for Elqui, red circles for Gaia-Enceladus-Sausage (GES), yellow circles for the Helmi Streams (HS), purple circles for the low energy group (low-E), green circles for the retrogrades (retro), blue circles for Sagittarius (Sag), green triangles for Typhon, black circles for the unassociated. The dashed line at $L_z = 0$ guides the eye in dividing the prograde and retrograde sides.}
    \label{fig:etotlz}
\end{figure}

\section{Results}\label{results}

\begin{table*}
  \centering
 \caption{Identification of globular cluster progenitors based on the dynamical analysis carried on in this paper.}\label{tab1}
  \begin{tabular}{| c c | c c | c c | c c |}
 \hline
 Cluster & Progenitor & Cluster & Progenitor & Cluster & Progenitor & Cluster & Progenitor \\
 \hline
 2MASS-GC01 & Disc & NGC 4590 & HS & NGC 6362 & Disc & NGC 6864 & GES \\
 2MASS-GC02 & Disc & NGC 4833 & GES & NGC 6366 & Disc & NGC 6934 & Unk \\
 AM 1 & Elqui & NGC 5024 & HS & NGC 6380 & Bulge & NGC 6981 & GES \\
 AM 4 & Unk & NGC 5053 & HS & NGC 6388 & low-E & NGC 7006 & Retro \\
 Arp 2 & Sag & NGC 5139 & Retro/Thamnos & NGC 6397 & Disc & NGC 7078 & Disc \\
 BH 140 & Unk & NGC 5272 & HS & NGC 6401 & Bulge & NGC 7089 & GES \\
 BH 261 & Disc & NGC 5286 & GES & NGC 6402 & low-E & NGC 7099 & GES \\
 Crater & Unk & NGC 5466 & Retro & NGC 6426 & HS & NGC 7492 & GES \\
 Djor 1 & Unk & NGC 5634 & HS & NGC 6440 & Bulge & Pal 1 & Aleph \\
 Djor 2 & Bulge & NGC 5694 & Retro & NGC 6441 & Disc & Pal 2 & GES \\
 E 3 & Aleph & NGC 5824 & HS & NGC 6453 & Bulge & Pal 3 & Cetus \\
 Eridanus & Sag & NGC 5897 & Unk & NGC 6496 & Disc & Pal 4 & Cetus \\
 ESO 93-08 & Aleph/Disc & NGC 5904 & GES & NGC 6517 & Bulge & Pal 5 & HS \\
 ESO 280-SC06 & GES & NGC 5927 & Disc & NGC 6522 & Bulge & Pal 6 & Bulge \\
 ESO 452-C11 & Bulge & NGC 5946 & low-E & NGC 6528 & Bulge & Pal 8 & Disc \\
 FSR 1713 & Disc & NGC 5986 & low-E & NGC 6535 & low-E & Pal 10 & Aleph/Disc \\
 FSR 1735 & Disc & NGC 6093 & low-E & NGC 6539 & Disc & Pal 11 & Disc \\
 FSR 1758 & Retro/Thamnos & NGC 6101 & Retro/Antaeus & NGC 6540 & Bulge & Pal 12 & Sag \\
 Gran 1 & Bulge & NGC 6121 & GES & NGC 6541 & Disc & Pal 13 & Retro \\
 Gran 2 & Retro/Thamnos & NGC 6139 & Disc & NGC 6544 & low-E & Pal 14 & Typhon \\
 Gran 3 & Unk & NGC 6155 & low-E & NGC 6553 & Disc & Pal 15 & Typhon \\
 Gran 4 & HS & NGC 6171 & Disc & NGC 6558 & Bulge & Patchick 122 & Disc \\
 Gran 5 & Bulge & NGC 6205 & GES & NGC 6569 & Disc & Patchick 126 & Disc \\
 HP 1 & Bulge & NGC 6218 & Disc & NGC 6584 & GES/HS & Pyxis & Elqui \\
 IC 1257 & GES & NGC 6229 & GES & NGC 6624 & Bulge & Rup 106 & HS \\
 IC 1276 & Disc & NGC 6235 & Disc & NGC 6626 & Bulge & Sag II & Unk \\
 IC 4499 & Retro & NGC 6254 & Disc & NGC 6637 & Bulge & Ter 1 & Bulge \\
 Laevens 3 & Sag & NGC 6256 & Bulge & NGC 6638 & Bulge & Ter 2 & Bulge \\
 Liller 1 & Bulge & NGC 6266 & Bulge & NGC 6642 & Bulge & Ter 3 & Disc \\
 Lynga 7 & Disc & NGC 6273 & low-E & NGC 6652 & Bulge & Ter 4 & Bulge \\
 NGC 104 & Disc & NGC 6284 & Unk & NGC 6656 & Disc & Ter 6 & Bulge \\
 NGC 288 & Retro & NGC 6287 & low-E & NGC 6681 & low-E & Ter 7 & Sag \\
 NGC 362 & GES & NGC 6293 & Bulge & NGC 6712 & low-E & Ter 8 & Sag \\
 NGC 1261 & GES & NGC 6304 & Disc & NGC 6715 & Sag & Ter 9 & Bulge \\
 NGC 1851 & GES & NGC 6316 & Disc & NGC 6717 & Bulge & Ter 10 & low-E \\
 NGC 1904 & GES & NGC 6325 & Bulge & NGC 6723 & Unk & Ter 12 & Disc \\
 NGC 2298 & GES & NGC 6333 & low-E & NGC 6749 & Disc & Ton 2 & Disc \\
 NGC 2419 & Sag & NGC 6341 & GES & NGC 6752 & Disc & UKS-1 & GES \\
 NGC 2808 & GES & NGC 6342 & Bulge & NGC 6760 & Disc & VVV-CL001 & Bulge \\
 NGC 3201 & Retro/Antaeus & NGC 6352 & Disc & NGC 6779 & GES & VVV-CL002 & Bulge \\
 NGC 4147 & GES & NGC 6355 & Bulge & NGC 6809 & low-E & VVV-CL160 & GES \\
 NGC 4372 & Disc & NGC 6356 & Disc & NGC 6838 & Disc & Whiting 1 & Sag \\
 \hline
 \end{tabular}
\tablefoot{The progenitors are the following: Aleph, Antaeus, bulge, Cetus, disc, Elqui, Gaia-Enceladus-Sausage (GES), Helmi Streams (HS), low energy group (low-E), Sagittarius (Sag), retrograde family (retro), Thamnos, Typhon, and unknown (Unk).}
\end{table*}

Recent studies on the non-conservation of the IoMs in the MW \citep[i.e.][]{2023A&A...673A..86P, 2025A&A...700A.240W, 2025MNRAS.542.1331D} due to the mergers, secular evolution of the Galaxy, effect of the rotating bar and a few other factors put into question the accuracy of dynamical analysis. This is the principal reason why, instead of focusing on only the IoMs and/or adiabatic invariants, this work presents a full sample of dynamical spaces (6 spaces for 10 different parameters) that helps in characterising the various progenitors. For each one of the progenitors, due to differences in the way the accretion event happened, its age, the mass of the accreted structure, and many other factors, there will be different signatures in different parameter spaces. The ability to explore a vast, multidimensional parameter space is fundamental in highlighting features and trends that allow to discriminate between the numerous orbital families present in our Galaxy.

\begin{figure}
\centering
\includegraphics[width=\hsize]{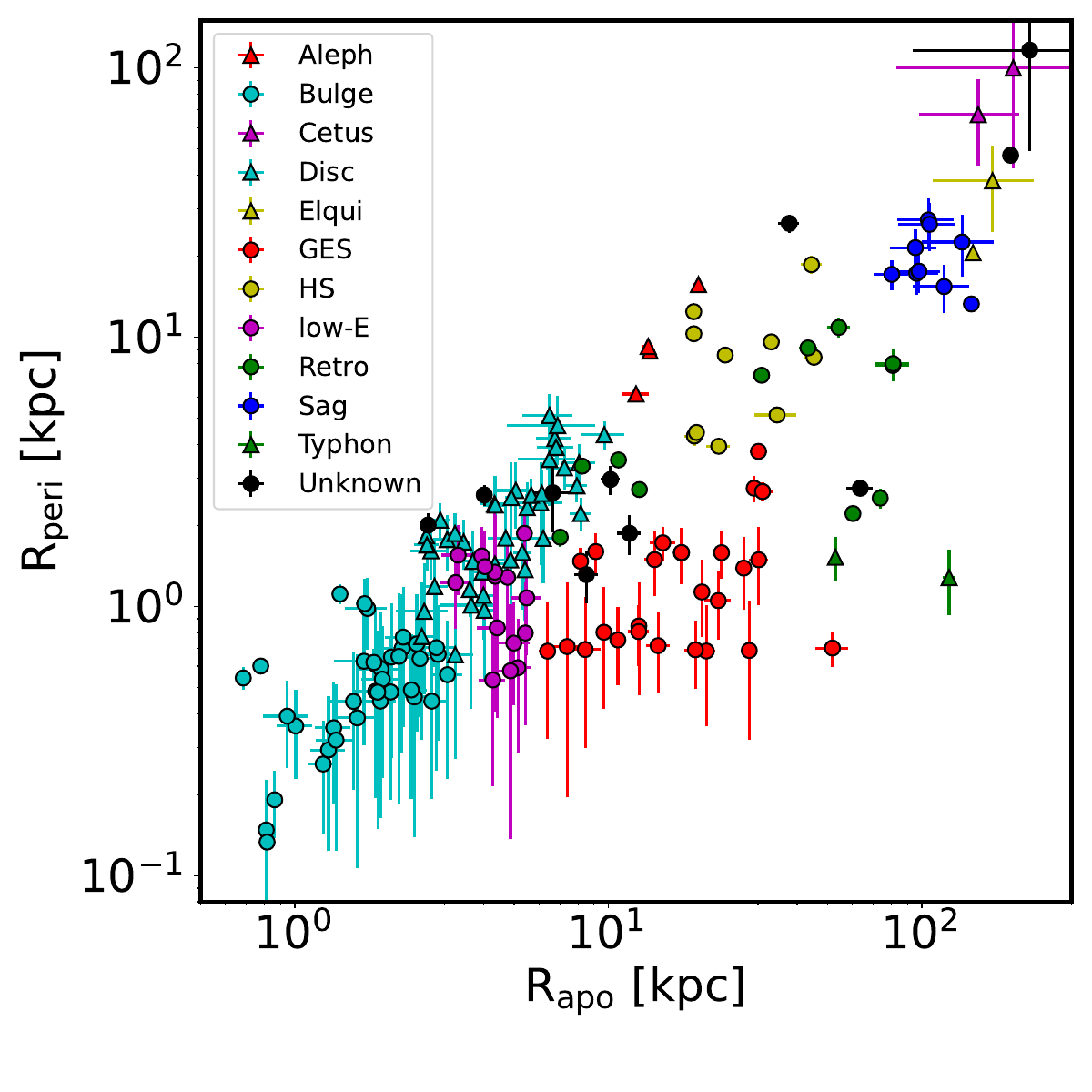}
\caption{The GC sample in the apocentre versus pericentre plane, in logarithmic scale. The uncertainties are computed as detailed in Sec.~\ref{out}. The GCs are marked depending on their identified progenitor as in Fig.~\ref{fig:etotlz}.}
    \label{fig:apoperi}
\end{figure}

The main results of our orbital analysis are summarised in Table~\ref{tab1}, where we list the GCs studied and their most likely progenitor. Figure~\ref{fig:etotlz} gives a quick overview of the composition of the orbital families in the commonly used $L_z-E_{tot}$ space. Throughout this section we will point out the main signatures in the various parameter spaces of the most important substructures inside the MW and provide boundaries for the values of the dynamical parameters deemed most important in assigning the GC to a given progenitor. To decide if a GC belongs to one family or another we check which \textquote{locus} it tracks across the most dynamical parameter spaces (ideally across all). Here and in the rest of the paper we refer to the locus of a substructure as the region of dynamical space populated by bona fide traces belonging to it. Some parameter spaces don't explicitly allow to distinguish between prograde and retrograde tracers, but that distinction is often useful to solve apparent overlapping between families (i.e. in Fig.~\ref{fig:apoperi} the green circles of retrograde GCs and the yellow circles of Helmi Streams GCs overlap but the two families are clearly discriminated by their sign of $L_z$). In Sec.~\ref{discu} we will cover the edge cases of GCs with the most uncertain associations.

\subsection{Structures considered and absent progenitors}\label{absentees}

This work is focused on the following structures: disc and bulge of the MW (making up the in-situ component), Gaia-Enceladus-Sausage \citep[GES,][]{2018ApJ...860L..11K, 2018MNRAS.478..611B, 2018ApJ...863..113H, 2018Natur.563...85H}, Sagittarius \citep[Sag,][]{1994Natur.370..194I, 2003ApJ...599.1082M}, the Helmi Streams \citep[HS,][]{1999Natur.402...53H, 2019A&A...625A...5K}, the retrograde halo (with a detailed discussion in Sec.~\ref{retro-disc}), the low energy group \citep[low-E, first labeled by][with a detailed discussion in Sec.~\ref{outliers}]{2019A&A...630L...4M}, Aleph \citep{2020ApJ...901...48N}, Cetus \citep{2009ApJ...700L..61N, 2019ApJ...881..164Y, 2022ApJ...930..103Y}, Typhon \citep{2022ApJ...935L..22T}, and Elqui \citep{2018ApJ...862..114S, 2019ApJ...885....3S}.

In our analysis we have seemingly ignored some known substructures, either because the progenitor locus substantially overlaps with well known structures or because it was not possible (due to lack or scarcity of data) to properly identify the progenitor locus. In both instances, this made it impossible to dynamically distinguish the newer candidates from the well established structures. For LMS-1/Wukong \citep{2020ApJ...901...48N, 2020ApJ...898L..37Y}, the stellar component identified by \citet{2024yCat..19200051M} overlaps with the loci of the HS and GES progenitors in all parameter spaces. The loci identified for Shakti and Shiva \citep{2024ApJ...964..104M} substantially overlap with the disc and low-E groups, while it has not been possible to produce a suitable dataset to identify the loci of Icarus \citep{2021ApJ...907L..16R, 2024ApJ...977..278R} and Pontus \citep{2022ApJ...926..107M, 2022ApJ...930L...9M}.
In the case of Nyx \citep{2020NatAs...4.1078N, 2020ApJ...903...25N, 2022NatAs...6..866N}, we don't include it in our analysis because of the mounting evidence that it is not an independent merger event \citep{2023MNRAS.520.5671H, 2023ApJ...955..129W}.

\subsection{Bulge}\label{bulge}
The MW Bulge GC population is characterised by having the least energetic GCs (cyan circles in Fig.~\ref{fig:etotlz}). While all types of orbits (prograde, retrograde or radial) are possible, the value of $\lvert L_z \rvert$ stays close to zero for all members of this group as they reside in the deepest part of the MW potential well. Likewise, having orbits largely confined to the bulge, these GCs occupy the bottom left corner of $R_{apo}-R_{peri}$ space (Fig.~\ref{fig:apoperi}). We consider as bulge GCs those with:
\begin{itemize}
  \item $E_{tot} < -1.75\times10^5 {\rm \ km^2 / s^2}$
  \item $R_{peri} < 1.5 {\rm \ kpc}$
  \item $R_{apo} < 3.5 {\rm \ kpc}$
\end{itemize}

\subsection{Disc}\label{disc}
While there are subtle kinematical differences between the populations of tracers inhabiting the thin and the thick discs of the MW, their dynamical distributions largely overlap and the two structures are better separated through chemical analysis \citep[i.e.][]{2014A&A...562A..71B, 2015ApJ...808..132H}. For this reason, in the scope of our present work, we refer to both thin and thick disc populations as belonging to one single structure.
The MW disc is quite prominently visible in the prograde portion of $L_z$-$E_{tot}$ space (cyan triangles in Fig.~\ref{fig:etotlz}). Consequently, the main features of disc GCs are the $\varepsilon \approx J_{\parallel} \leq -0.5$ (usually two very similar values, as they both gauge the rotational support of an orbit) and $ecc \leq 0.6$. Disc GCs are also well identified in the most prograde sector of $J_{\parallel}-J_{\perp}$ space (Fig.~\ref{fig:jparaperp}) and trace a characteristic sequence in the $R_{peri}-ecc$ space (Fig.~\ref{fig:periecc}). Disc GCs generally have the following defining characteristics:
\begin{itemize}
  \item $-2.0\times10^5 {\rm \ km^2 / s^2} < E_{tot} < -1.1\times10^5 {\rm \ km^2 / s^2}$
  \item $L_z > 0 {\rm \ kpc \ km / s}$
  \item $0.6 {\rm \ kpc} < R_{peri} < 6 {\rm \ kpc}$
  \item $Ecc < 0.7$
  \item $J_{\parallel}/\varepsilon < -0.4$
\end{itemize}

\subsection{Gaia-Enceladus-Sausage (GES)}\label{GES}

As the most massive accretion event of our Galaxy, the GES is prominently visible as a central, mildly energetic overdensity in $Lz$-$E_{tot}$ space (red circles in Fig.~\ref{fig:etotlz}). The population of GCs associated with this event is characterised by radial orbits (Fig.~\ref{fig:jparaperp}) with high $ecc$ at relatively low $R_{peri}$ (Fig.~\ref{fig:periecc}). GES GCs are characterised by:
\begin{itemize}
  \item $-1.5\times10^5 {\rm \ km^2 / s^2} < E_{tot} < -0.6\times10^5 {\rm \ km^2 / s^2}$
  \item $Ecc > 0.7$
  \item $ 4 {\rm \ kpc} < Z_{max} < 40 {\rm \ kpc}$
  \item $-0.3 < J_{\parallel}/\varepsilon < 0.3$
  \item $J_\perp < 0.0$
\end{itemize}

\subsection{Sagittarius (Sag)}\label{sag}
The ongoing merger of the Sagittarius dwarf galaxy is the most recent accretion event to affect our Galaxy, and it is still ongoing and visible in the sky in the form of a wide, long stellar stream.
The GCs associated with the Sag event exhibit highly energetic, prograde orbits (blue circles in Fig.~\ref{fig:etotlz}), are among those with the highest $R_{peri}$ and $R_{apo}$ (Fig.~\ref{fig:apoperi}) and also have high $L_{\perp}$ (Fig.~\ref{fig:lperplz}). We define as Sag GCs those with:
\begin{itemize}
    \item $E_{tot} > -0.5\times10^5 {\rm \ km^2 / s^2}$
    \item $-3.0\times10^3 {\rm \ kpc \ km/s} < L_z < -0.5\times10^3 {\rm \ kpc \ km/s}$
    \item $4.0\times10^3 {\rm \ kpc \ km/s} < L_{\perp} < 7\times10^3 {\rm \ kpc \ km/s}$
\end{itemize}

\subsection{Helmi Streams (HS)}\label{Helmi}
The progenitor of the Helmi Streams has been one of the first accreted substructures identified through the study of IoMs. The population belonging to this accretion event is characterised by prograde, high-energy orbits between the disc and the Sag populations in $L_z-E_{tot}$ space (yellow circles in Fig.~\ref{fig:etotlz}). The HS GCs also occupy a well isolated portion of the $L_z-L_{\perp}$ plane (Fig.~\ref{fig:lperplz}) and a distinctive sequence in $R_{peri}-ecc$ space (Fig.~\ref{fig:periecc}). Our selection for the HS GCs is:
\begin{itemize}
  \item $-1.0\times10^5 {\rm \ km^2 / s^2} < E_{tot} < -0.5\times10^5 {\rm \ km^2 / s^2}$
  \item $L_z < -0.5\times10^3 {\rm \ kpc \ km/s}$
  \item $1.0\times10^3 {\rm \ kpc \ km/s} < L_{\perp} < 4.0\times10^3 {\rm \ kpc \ km/s}$
\end{itemize}

\subsection{Retrogrades}\label{retro}
The full characterisation of the retrograde portion of the halo is an ongoing effort that has seen several candidate merger events proposed over recent years. Since some of the candidate progenitors have been discarded, the consensus around others is still ongoing and new ones are being proposed, we refer to Sec.~\ref{retro-disc} for a more in-depth discussion of each progenitor and here we limit ourselves to sketching defining traits of the whole retrograde family.
The retrograde GCs can be distinguished clearly in $L_z-L_{\perp}$ space (Fig.~\ref{fig:lperplz}) and populate a distinctive diagonal streak of increasing $\varepsilon$ for decreasing $\lvert Z_{max} \rvert$ (Fig.~\ref{fig:circzmax}). We identify as members of this group the GCs having:
\begin{itemize}
  \item $E_{tot} > -1.5\times10^5 {\rm \ km^2 / s^2}$
  \item $L_z > 0.5\times10^3 {\rm \ kpc \ km/s}$
  \item $Z_{max} > 4 {\rm \ kpc}$
\end{itemize}

\begin{figure}
\centering
\includegraphics[width=\hsize]{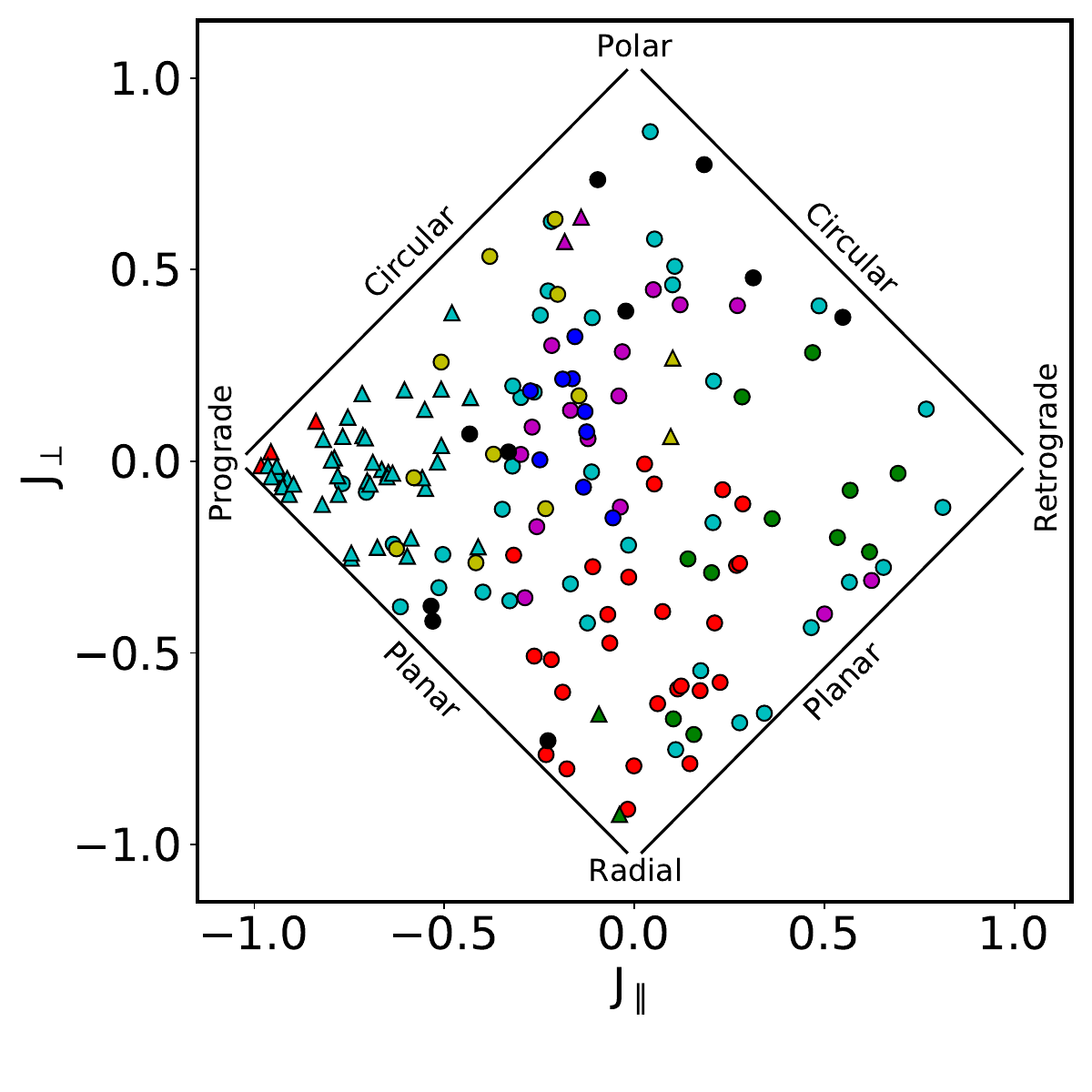}
\caption{The GC sample in the parallel versus perpendicular action space. GCs with $J_{\parallel} \approx -1$ are on nearly circular prograde orbits, conversely $J_{\parallel} \approx 1$ identifies nearly circular retrograde orbits. GCs on radial orbits have $J_{\perp} \approx -1$ while those on polar orbits have $J_{\perp} \approx 1$. The GCs are marked depending on their identified progenitor as in Fig.~\ref{fig:etotlz}.}
    \label{fig:jparaperp}
\end{figure}

\begin{figure}
\centering
\includegraphics[width=\hsize]{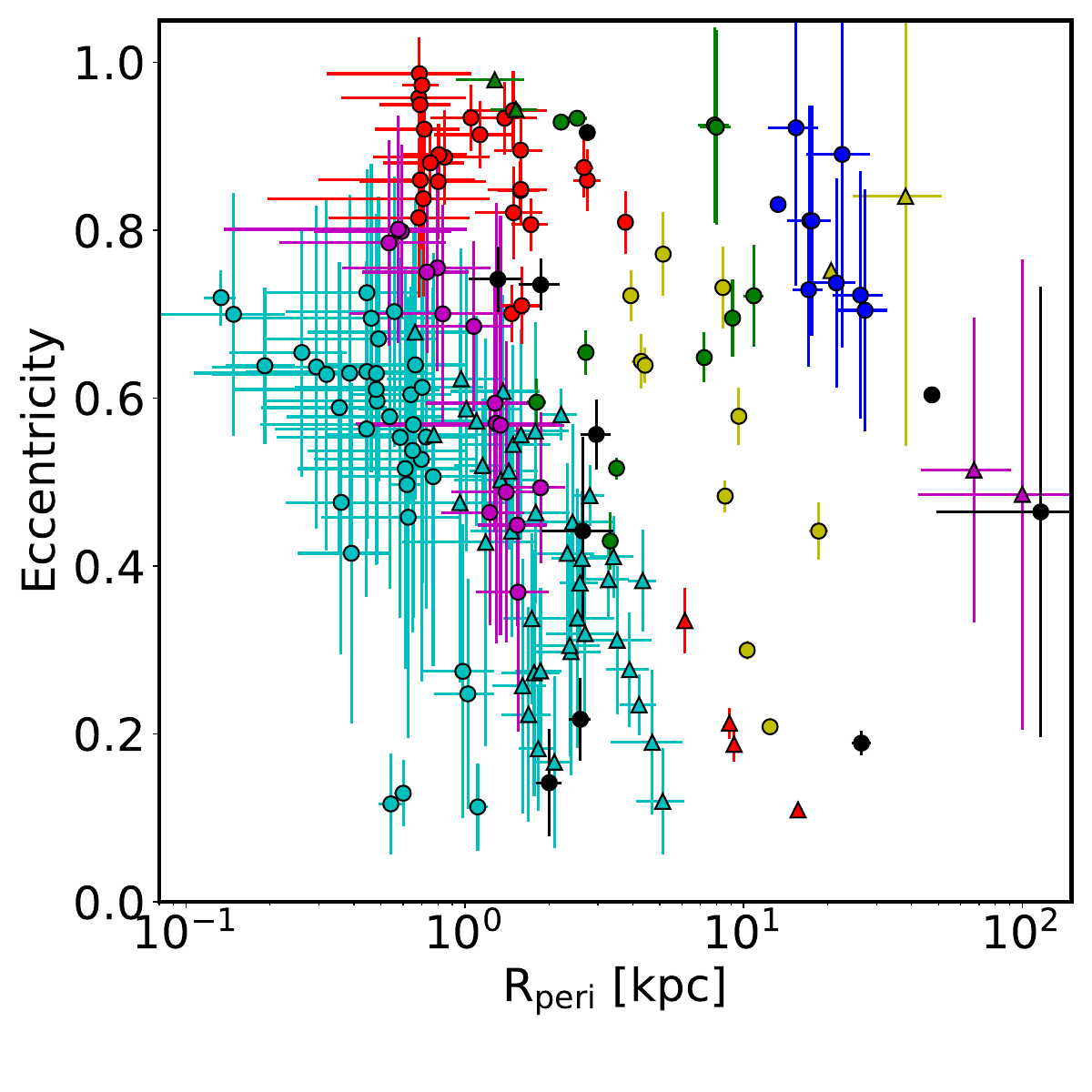}
\caption{The GC sample in the pericentre versus eccentricity plane, in semilogarithmic scale. The uncertainties are computed as detailed in Sec.~\ref{out}. The GCs are marked depending on their identified progenitor as in Fig.~\ref{fig:etotlz}.}
    \label{fig:periecc}
\end{figure}

\subsection{Low-energy group (low-E)}\label{lowE}
This group has been associated to various different accretion events and in-situ structures and for the moment we will keep this general label.
The GCs associated with this group are characterised by energies lower than the GES but higher than the bulge and small values of $\lvert L_z \lvert$ (purple circles in Fig.~\ref{fig:etotlz}). Similarly, the GCs in the low-E group populate a region of mid $\varepsilon$ and $Z_{max}$ (Fig.~\ref{fig:circzmax}). Low-E GCs generally have:
\begin{itemize} 
  \item $-1.8\times10^5 {\rm \ km^2 / s^2} < E_{tot} -1.4\times10^5 {\rm \ km^2 / s^2}$
  \item $0.5 {\rm \ kpc} < R_{peri} < 2 {\rm \ kpc}$
  \item $3 {\rm \ kpc} < R_{apo} < 6 {\rm \ kpc}$
  \item $3 {\rm \ kpc} < Z_{max} < 6 {\rm \ kpc}$
  \item $-0.5 < \varepsilon < 0.5$
\end{itemize}

\subsection{Other structures}\label{others}
As already pointed out, the wealth of data that became available roughly in the last decade has allowed to populate much more the different dynamical parameter spaces and highlight smaller confirmed merger events (and candidates). The following progenitors have only a small number of GCs associated to each of them.

\subsubsection{Aleph}\label{aleph}
This structure appears to be a direct continuation of the disc at high energies with some of the most prograde GCs (red triangles in Fig.~\ref{fig:etotlz}). The GCs belonging to this progenitor also exhibit higher $R_{peri} \ {\rm and} \ R_{apo}$ than their counterparts in the disc (Fig.~\ref{fig:apoperi}). Alpeh GCs have:
\begin{itemize} 
  \item $E_{tot} > -1.1\times10^5 {\rm \ km^2 / s^2}$
  \item $R_{peri} > 5 {\rm \ kpc}$
  \item $R_{apo} > 10 {\rm \ kpc}$
  \item $Ecc < 0.4$
  \item $L_z < -1.75\times10^3 {\rm \ kpc \ km/s}$
\end{itemize}

\subsubsection{Cetus}\label{cetus}
This merger, identified as a stream on a very polar orbit, is composed of prograde and very energetic tracers, very close to Sag in $L_z-E_{tot}$ space (purple triangles in Fig.~\ref{fig:etotlz}). It is possible to distinguish the genuine members of the Cetus merger for their extremely high $R_{peri}$, $R_{apo}$ (Fig.~\ref{fig:apoperi}), $L_{\perp}$ (Fig.~\ref{fig:lperplz}) and almost fully polar orbits (Fig.~\ref{fig:jparaperp}). The GCs belonging to Cetus have:
\begin{itemize} 
  \item $E_{tot} > -0.5\times10^5 {\rm \ km^2 / s^2}$
  \item $L_z \approx -2.0\times10^3 {\rm \ kpc \ km/s}$
  \item $L_{\perp} > 11.0\times10^3 {\rm \ kpc \ km/s}$
\end{itemize}

\subsubsection{Typhon}\label{typhon}
This is another merger discovered as a surviving stream with a long, radial orbit. We tentatively associate two high energy GCs to this progenitor. These GCs partially overlap in some parameter spaces with the Typhon stellar populations identified by \citet{2022ApJ...935L..22T} and \citet{2023A&A...670L...2D} and are characterised by radial orbits (green triangles in Fig.~\ref{fig:jparaperp}) and very high $R_{apo}$ coupled to very low $R_{peri}$ (Fig.~\ref{fig:apoperi}). We tentatively identify as Typhon GCs those having:
\begin{itemize} 
  \item $E_{tot} > -0.7\times10^5 {\rm \ km^2 / s^2}$
  \item $1.0 {\rm \ kpc} < R_{peri} < 2.0 {\rm \ kpc}$
  \item $R_{apo} > 50 {\rm \ kpc}$
  \item $Z_{max} > 40 {\rm \ kpc}$ 
\end{itemize}

\subsubsection{Elqui}\label{elqui}
\begin{figure}
\centering
\includegraphics[width=\hsize]{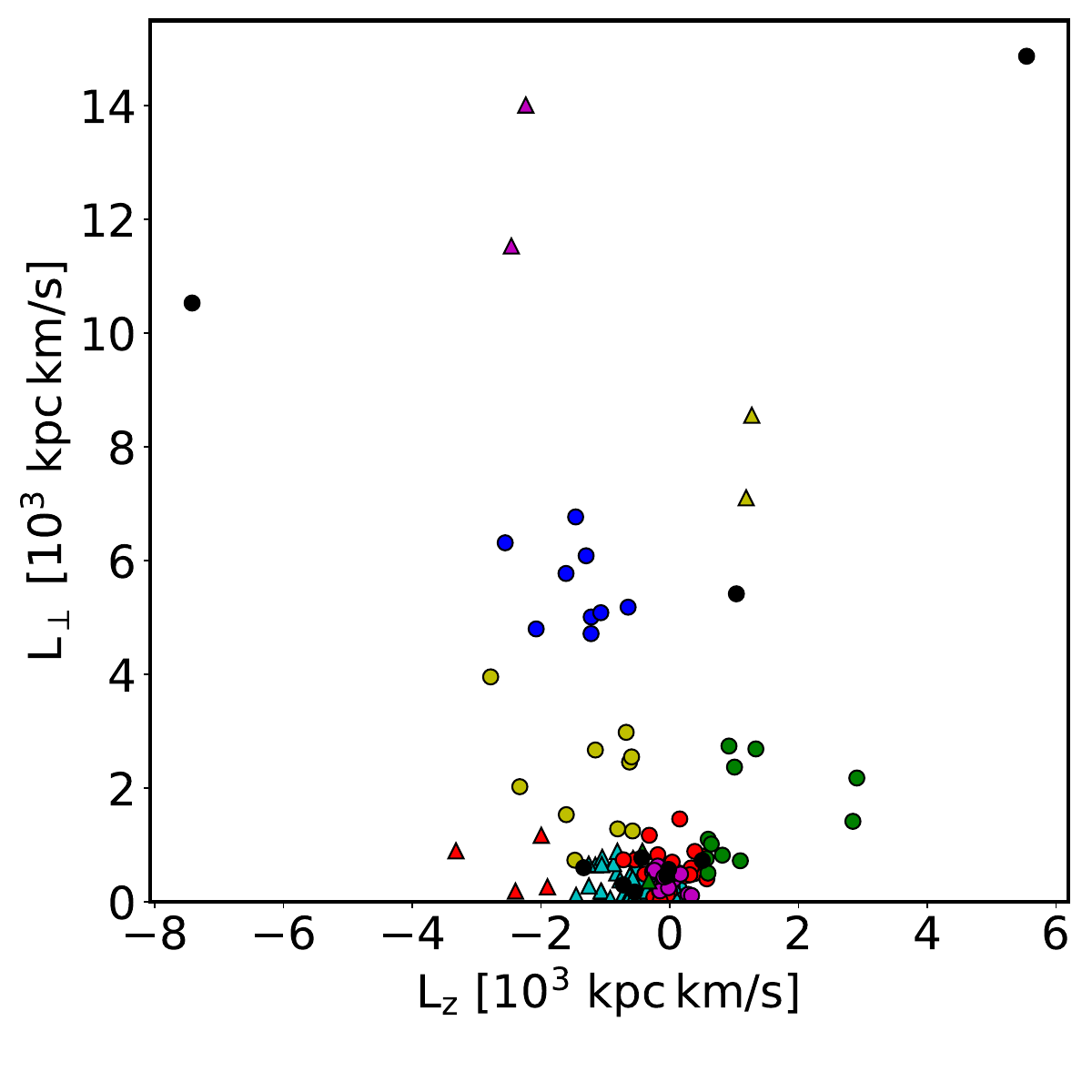}
\caption{The GC sample in the vertical versus perpendicular angular momentum plane. As clear by this figure, the majority of GCs are crowded in the low $L_{\perp}$ region, with the loci of different progenitors overlapping, but this space is very helpful to distinguish the various high energy families. The GCs are marked depending on their identified progenitor as in Fig.~\ref{fig:etotlz}.}
    \label{fig:lperplz}
\end{figure}

The stream of debris produced by this merger is one of the most distant identified \citep{2018ApJ...862..114S, 2019ApJ...885....3S}. The associated GCs are retrograde and at high energies (yellow triangles in Fig.~\ref{fig:etotlz}) but they have much higher $R_{peri}$, $R_{apo}$ (Fig.~\ref{fig:apoperi}), and $L_{\perp}$ (Fig.~\ref{fig:lperplz}) with respect to the retrograde group. Being quite isolated from the rest of the retrograde family in almost all dynamical spaces, we are confident in assigning two GCs directly to Elqui. These GCs have:
\begin{itemize}
    \item $E_{tot} > -0.5\times10^5 {\rm \ km^2 / s^2}$
    \item $L_z \approx 1.0\times10^3 {\rm \ kpc \ km/s}$
    \item $7.0\times10^3 {\rm \ kpc \ km/s} < L_{\perp} < 9.0\times10^3 {\rm \ kpc \ km/s}$
\end{itemize}

\begin{figure}
\centering
\includegraphics[width=\hsize]{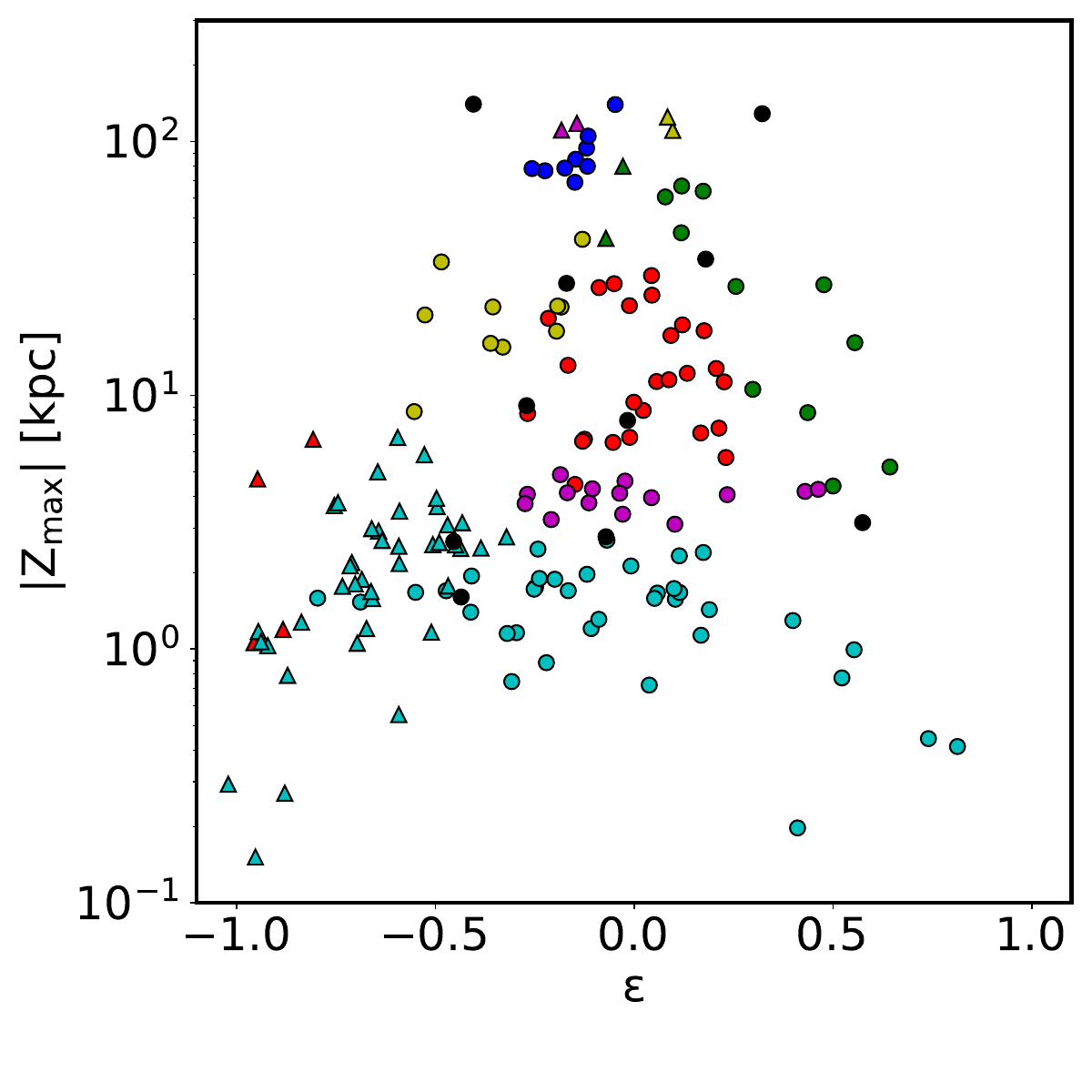}
\caption{The GC sample in the circularity versus maximum vertical excursion plane, in semilogarithmic scale. The GCs are marked depending on their identified progenitor as in Fig.~\ref{fig:etotlz}.}
    \label{fig:circzmax}
\end{figure}

\section{Discussion}\label{discu}

The main difference between this work and most in the literature is the presence of the rotating bar of the MW in the model underlying the orbital integration. We tested the direct effect of including/excluding the bar by running \textsc{OrbIT} with and without the rotating bar and noting the differences in the outputs. The main effect of the bar is in changing pericentres and eccentricities for GCs roughly below $E_{tot} = -1.7\times10^5 {\rm \ km^2/s^2}$ and for some of those with highly radial orbits (i.e. very small pericentres). This translates into some changes in the classification of the GCs in the lowest energy groups (bulge, disc, GES and low-energy) but leaves basically untouched the classification of the GCs belonging to the other families. Without the bar Gran 5, NGC 6517, NGC 6626 (all previously classified bulge GCs), NGC 6121 (previously GES), NGC 6723 (previously unclassified), and NGC 6171 (previously disc) would be classified as low-energy GCs. Three clusters (NGC 6256, NGC 6266 and NGC 6540) would move from the bulge to the disc family and one (2MASS-GC02) would move from the disc to the bulge family. Finally, NGC 6284 would be identified as a GES member. These changes in associations of GCs and progenitors make our whole classification much closer to literature ones \citep[i.e.][]{2019A&A...630L...4M} because our model without a bar is essentially a variant of the widely used \citet{2017MNRAS.465...76M} model.

\subsection{Outliers within the groups}\label{outliers}

While we identified defining characteristics for each family, some members do stand out thanks to peculiar properties. In the following, we cover these and the GCs contended between groups.

Starting from the bulge, it is important to mention the small number of GCs with $\varepsilon$ and $J_{\parallel}$ values suggesting highly counter-rotating orbits (the cyan circles at the bottom right of Fig.~\ref{fig:circzmax} with $\varepsilon > 0.4$). VVV-CL001, Djor 2, Terzan 2, Terzan 6, Liller 1, and NGC 6325 appear to be clearly separated from the main bulge population by their peculiar kinematics. Tests show that the spherical component of the bulge is partly responsible for their extreme kinematics, but even removing that component these GCs remain on clearly counter-rotating orbits (just not as extreme). Gran 1 stands out for another reason, as one of the least eccentric GCs of the entire sample, with a very polar orbit confined to the inner bulge (the topmost cyan circle at $J_{\perp} \approx 1.0$ in Fig.~\ref{fig:jparaperp}).

At the \textquote{interface} between the bulge and the disc groupings there are four GCs with quite circular orbits (cyan circles with $J_{\parallel} \geq 0.6$ in Fig.~\ref{fig:jparaperp}). Despite this, NGC 6256, NGC 6266, NGC 6540, and Terzan 1 reside deep into the potential well, with very small $R_{apo}$ and $R_{peri}$, so we assigned them to the bulge.
Conversely, 2MASS-GC02 is the least energetic of the disc GCs but its $R_{apo}$ and $R_{peri}$ are not low enough to make it a resident of the bulge region, so we assigned it to the disc family. Four out of five of these GCs are exactly the ones moving between the bulge and disc families depending on the presence/absence of the rotating bar (Sec.~\ref{discu}) and show the influence of different choices of bulge models.

Moving on to the low-E group, it is worth reminding that it has been the home of tracers that couldn't be unambiguously attributed to either the GES or the MW bulge. Several authors linked some or all members of this group to different accretion events, Kraken \citep{2019MNRAS.486.3180K, 2020MNRAS.498.2472K}, Koala \citep{2020MNRAS.493..847F}, Heracles \citep{2021MNRAS.500.1385H}, which might possibly be the same structure. Other works highlighted in-situ populations with similar chemo-kinematical signatures, Aurora \citep{2022MNRAS.514..689B, 2022ApJ...938...21M}, Poor Old Heart \citep{2022ApJ...941...45R}, arguing that all these structures (accreted and in-situ) are the different facets of the same stellar population constituting the proto-Galaxy. Nevertheless, due to their quite tight distribution in dynamical parameter spaces, we have kept these clusters grouped together. The only two significant outliers in this family are NGC 6388 and NGC 6535 being the most retrograde and planar low-E GCs (they are the rightmost purple circles in Fig.~\ref{fig:circzmax}). Both GCs, while very retrograde, don't have enough energy to connect with the retrograde family or with the hypothesised Thamnos members (discussed in Sec.~\ref{retro-disc}). Interestingly, they remain quite close in all parameter spaces, which might hint at a shared history.

The GES group of GCs has several slight outliers. Pal 2 is one of the most energetic GES GCs with a very radial orbit, a very low $R_{peri}$ and very high $R_{apo}$ (rightmost red circle in Fig.~\ref{fig:apoperi}) and $\lvert Z_{max} \rvert$ but all the other parameters identify it as a belonging to the GES progenitor. NGC 5904 and NGC 7492 are compatible with GES in most parameter spaces but have the highest $L_{\perp}$ and $R_{peri}$ of the family. Similarly, the high $ecc$ and radial orbit of NGC 6584 identify it as a GES GC even if it has very high $\lvert Z_{max} \rvert \ {\rm and} \ R_{peri}$ (these three GCs are the highest red circles in Fig.~\ref{fig:apoperi}). NGC 6205 and NGC 7099 fit quite well amid the GES family except for their uncharacteristically low $ecc$ (red circles at $ecc \approx 0.7$ in Fig.~\ref{fig:periecc}). Given the nature of GES as the most massive merger event of the MW, it is quite possible for its dynamical signature to not be as compact as other smaller and more recent merger events (i.e. Sag) and present a low number of outliers (6 out of 26 GCs).

Moving towards prograde and higher energy space, ESO 93-08 and Pal 10 are the two uncertain attributions to the Aleph structure (lowest red triangles in Fig.~\ref{fig:circzmax}). While having higher $E_{tot} \ {\rm and} \ \lvert L_z \rvert$ than disc GCs, they have much lower $\lvert Z_{max} \rvert \ {\rm and} \ L_{\perp}$ than the other members of Aleph and therefore might be the very outermost disc GCs. Given the uncertainty in the nature and characteristics of the Aleph progenitor, it is unclear if these GCs have a tighter link to the disc or not.

NGC 4590, NGC 5824 and NGC 6426 are the GCs on the most prograde circular orbits for the HS family (yellow circles with $J_{\parallel}<-0.5$ in Fig.~\ref{fig:jparaperp}), similar to disc GCs, but they have $R_{apo}, \ Z_{max}, \ {\rm and} \ L_{\perp}$ in line with the HS group and they are on the HS sequence in $R_{peri}-ecc$ space. Leaving debris with quite prograde orbits is to be expected from a merger generally believed to be prograde (even if not planar) and quite massive at the time of accretion \citep{1999Natur.402...53H}.

The Sag group is well isolated in all parameter spaces with only Eridanus being slightly an outlier for its higher $\lvert Z_{max} \rvert$.

\subsection{The retrograde family}\label{retro-disc}
The retrograde halo has been populated by several candidate progenitors over the years, starting with Sequoia \citep{2018MNRAS.478.5449M, 2019ApJ...870L..24B, 2019ApJ...874L..35M, 2019MNRAS.488.1235M} and Elqui \citep{2018ApJ...862..114S, 2019ApJ...885....3S} to several more. Studies on simulations recreating the complex dynamics of galactic mergers suggest that high mass-ratio events spread their debris across large portions of parameter space, spilling copiously also in the retrograde portion of the halo, leading to these debris being mistaken for independent mergers \citep{2020A&A...642L..18K, 2022ApJ...937...12A, 2023A&A...677A..90K}. Arjuna \citep[proposed by ][]{2020ApJ...901...48N} has been rejected for this very reason and its sibling I'itoi (from the same work) is facing similar scrutiny, hampered by a dearth of chemical information \citep{2021ApJ...923...92N, 2023MNRAS.520.5671H}.

Thamnos \citep{2019A&A...631L...9K} is another candidate member of the retrograde family, first identified as a couple of overdensities very close to each other in the mid-energy, retrograde portion of $L_z-E_{tot}$ space. \citet{2019A&A...631L...9K} and subsequent works estimated a very low mass for the Thamnos progenitor ($M_{Thamnos} \leq 5 \cdot 10^7 M_{\odot}$), making it very unlikely it would harbour any GC capable of surviving the merging event while keeping a recognisable dynamical signature up to the present time. Moreover, some recent works on the chemical characterisation of substructures seem to indicate that the Thamnos stellar population is heavily contaminated by in-situ and GES tracers \citep[i.e. Fig. 15 from][]{2022A&A...665A..58R}.
Nevertheless, based on a purely dynamical analysis, there are several clusters unassociated with other progenitors or within the retrograde family that fall into or near the loci identified by the Thamnos stellar component in the various dynamical spaces. NGC 5139 (also known as $\omega$Cen) and FSR 1758 are the GCs that track more closely the Thamnos stellar population across all parameter spaces while Gran 2 exhibits a slightly out-of-plane, more polar orbit (with higher $\lvert Z_{max} \rvert$ and $L_{\perp}$). The aforementioned NGC 6388 and NGC 6535 have retrograde, planar orbits very similar to the Thamnos tracers but they reside much deeper in the MW potential well and have lower $E_{tot}, \ R_{apo} \ {\rm and} \ R_{peri}$.

Antaeus/L-RL64 is an extremely retrograde and highly energetic merger remnant identified by \citet{2022ApJ...936L...3O} and \citet{2022A&A...665A..58R}. More recently, the small structure of ED-3 has been identified in the retrograde halo \citep{2023A&A...670L...2D}, occupying loci connected or contiguous to Antaeus/L-RL64 in all parameter spaces. The association of the two structures has been corroborated by the findings of \citet{2024A&A...684A..37C}. Based on their heavily retrograde orbits, NGC 3201 and NGC 6101 could be associated with this structure. Even if the two clusters don't directly track the stars identified as Antaeus members by \citet{2022ApJ...936L...3O} and \citet{2022yCat..36709002D}, NGC 3201 follows quite closely the loci of the ED-3 members in all dynamical spaces, allowing us to establish a tentative association between the two GCs and the Antaeus progenitor.

\subsection{Unclassified GCs}\label{unknown}

Some MW GCs share some defining characteristics with the above mentioned groups but also appear to be very discordant in other parameter spaces. We will quickly review this list of outliers that cannot be unequivocally assigned to one of the progenitors.

AM-4 sits amid the retrograde group in $L_z-E_{tot}$ but has much higher $R_{peri}$ and much lower $ecc$ that distinguish it as a clear extreme outsider to this family. Crater is too extreme in all parameter spaces to be a part of any retrograde family. Sag II is Crater's counterpart in the prograde portion of dynamical space, being a distant outlier from all identified groups. NGC 6934 is too energetic for, and has an higher $R_{apo}$ than, Helmi Streams GCs but has $R_{peri}$ and $\lvert Z_{max} \rvert$ too low for Sag GCs and its orbit is too radial for both families. Sitting between the loci of the low-energy and GES families, NGC 6284 has $R_{apo}, \ R_{peri}, \ {\rm and} \ \lvert Z_{max} \rvert$ too high to belong to the former group and $ecc$ too low, coupled to an orbit more polar than radial, to be a GES GC. BH 140 and Djor 1 have many characteristics of disc GCs, with prograde, circular orbits, but their $R_{apo}, \ R_{peri}, \ {\rm and} \ ecc$ are too high compared to the other disc GCs. The orbit of NGC 5897 is not circular enough for it to be a (thick) disc GC and not radial enough to belong to the GES. Similarly, its $R_{apo}, \ R_{peri}, \ {\rm and} \ \lvert Z_{max} \rvert$ are too high for the disc and its $ecc$ is too low for the GES. Gran 3 has a retrograde, circular, very low $ecc$ orbit that sets it apart, at its energy range, from nearby identified groups. NGC 6723 similarly has a very low $ecc$ (so it is very improbable that it belongs to the low-E group) and is characterised by a peculiar, extremely polar orbit.

\subsection{Literature comparison}\label{lietrature}

Reviewing some recent literature on the classification of the MW GC population, we don't do a direct comparison with the work of \citet{2024A&A...687A.201B} as they used a selection base on ages and chemistry, while we used purely dynamical information. Similarly, the selection done by \citet{2024MNRAS.528.3198B}, while taking into account the dynamical component, is guided and calibrated using the [Al/Fe] abundance ratio and only distinguishes between the in-situ and accreted component. Nevertheless we generally agree on the classification by \citet{2024MNRAS.528.3198B}, with the notable exception of the low-E family \citep[whose members the authors mostly classify as in-situ, probably attributing them to the Aurora population, ][]{2022MNRAS.514..689B}. For the families identified in both works, we also find general agreement with the classification by \citet{2024OJAp....7E..23C}, again with the notable exception of the low-E group, identified as in-situ by these authors. Our results closely agree with the classification of the high- and mid-energy GCs of \citet{2022MNRAS.513.4107C, 2023RAA....23a5013S, 2025RNAAS...9...64M}, but we have different findings in the lowest energy groups (bulge, disc, low-E). All these works use an underlying \citet{2017MNRAS.465...76M} static potential and is clear that the discrepancies in the classification of the least energetic GCs depend on the impact of the rotating bar on the orbital parameters. In the case of \citet{2025RNAAS...9...64M} it is worth also noting that part of the differences are due to the authors not including Aleph in their list of progenitors and in us slightly unpacking their \textquote{high energy} group into some known progenitors (and Typhon). We note several differences with respect to the classification made by \citet{2022ApJ...926..107M} and the principal reason is the use of a partially different set of progenitors, namely LMS-1/Wukong, Pontus and an unknown retrograde merger (see Sec.~\ref{absentees} for our reasons to not include these progenitors in our list). One of the first works on the topic to use an underlying potential including a rotating bar is \citet{2020MNRAS.491.3251P} but they use a different classification (bulge, disc, inner halo, outer halo) and don't discriminate between the various merger events. Nevertheless our results closely agree in the list of bulge and disc clusters, with minor discrepancies due to the different cut in $ecc$ used to determine if a GC belongs to the disc or not. Another work including a rotating bar in the underlying potential is \citet{2023A&A...669A.136G}, working on a small subsample of clusters. Our results are in general agreement with theirs, with orbital parameters consistent within the errors.

\section{Conclusions}\label{conclusion}

We have presented \textsc{OrbIT}, a new in-house tool aimed at precise and efficient orbit integration for the study of the galactic dynamics. The novelty of the code resides in fully accounting for a time-varying potential at all stages, especially for the recovery of the orbital parameters and their associated errors. \textsc{OrbIT}, with its up-to-date potential model of the MW including the rotating bar and a spherical bulge component, has already been used to characterise the RR Lyrae population of the bulge \citep{2024A&A...687A.312O}. Here, we used \textsc{OrbIT} to study the dynamics of the galactic GCs and to classify them among the various families of merger events.

By using the information derived from six different dynamical parameter spaces (of orbital parameters, IoMs and adiabatic invariants), we provided a full dynamical characterisation of the in-situ population and of the different progenitors. The inclusion of the rotating bar changes the associations of low energy GCs while the full dynamical picture painted by our results allows for clear associations for a number of previously unassociated high energy GCs to various structures (4 GCs to Aleph and 2 each to Cetus, Typhon, Elqui and Antaeus).

We also proposed the association of 3 GCs to the Thamnos progenitor, further complicating the conundrum represented by this very small old merger.

A number of GCs remain of ambiguous ancestry or are clear outliers inside their families and some are still without any associated progenitor. This work is but a stepping stone in the process of thorough characterisation of the MW inventory and our purely dynamical view has to be complemented with cutting edge chronochemical studies \citep[i.e. the CARMA project, ][]{2023A&A...680A..20M, 2025arXiv250220436A, 2025arXiv250302939C}.

In future work we will explore in more depth the effect of the rotating bar, both with its direct gravitational pull affecting the central region of the MW and with the resonances it generates as a time-varying component of the potential.

\begin{acknowledgements}

The authors would like to thank the anonymous referee for insightful comments that helped improve and clarify the manuscript. Thanks also go to D. Massari, E. Ceccarelli, M. Bellazzini, A. Mucciarelli and E. Dodd for valuable discussions. MDL acknowledges financial support from the National Agency for Research and Development (ANID), Millenium Science Initiative, ICN12\_009 and from the project \textquote{LEGO – Reconstructing the building blocks of the Galaxy by chemical tagging} (PI: Mucciarelli) granted by the Italian MUR through contract PRIN2022LLP8TK\_001. MZ acknowledges support by the National Agency for Research and Development (ANID) Millenium Science Initiative, ICN12\_009 and AIM23-0001, awarded to the Millennium Institute of Astrophysics (MAS) and the ANID BASAL Center for Astrophysics and Associated Technologies (CATA) through grant FB210003, and from ANID FONDECYT Regular No. 1230731. BA-T acknowledges support from the ANID Doctoral Fellowship through grant number 21231305. FG gratefully acknowledges support from the French National Research Agency (ANR) funded projects \textquote{MWDisc} (ANR-20-CE31-0004) and \textquote{Pristine} (ANR-18-CE31-0017). This research made use of the Astropy \citep{2013A&A...558A..33A, 2018AJ....156..123A}, Matplotlib \citep{2007CSE.....9...90H} and Numpy \citep{harris2020array} packages.

\end{acknowledgements}

\section*{Data Availability}

The data underlying this article come from public sources listed in Section ~\ref{data}. Table ~\ref{tab2} is only available in electronic form at the CDS via anonymous ftp to cdsarc.u-strasbg.fr (130.79.128.5) or via http://cdsweb.u-strasbg.fr/cgi-bin/qcat?J/A+A/ (an excerpt of the table is provided in Appendix ~\ref{appb}).

\bibliographystyle{aa}
\bibliography{orbit}

\begin{appendix}

\section{Impact of observational errors: worst cases}\label{obserrs}

\begin{figure}[h]
\centering
\includegraphics[width=\hsize]{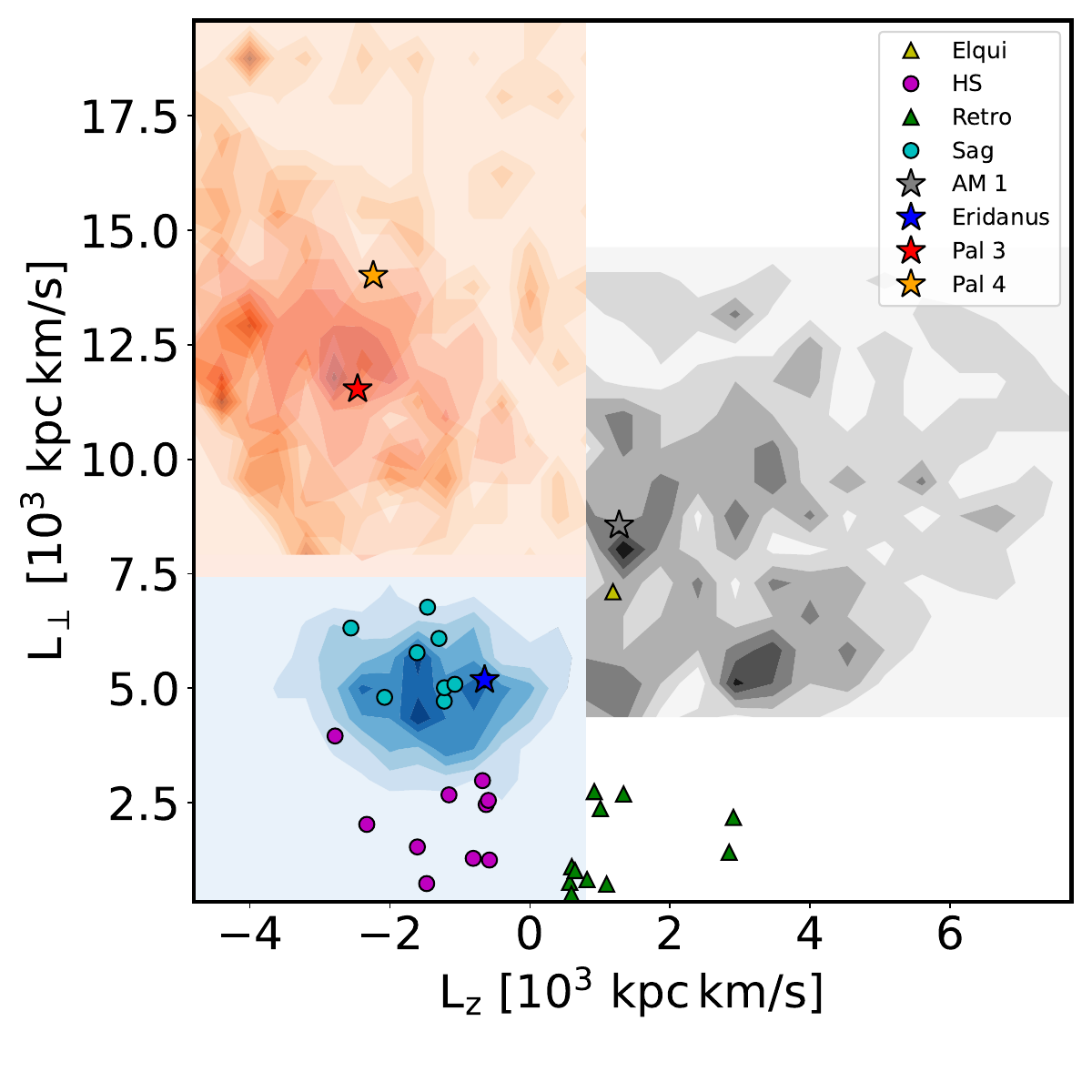}
\caption{$L_z-L_{\perp}$ space distribution of the GCs with the highest distance uncertainties (filled stars). The underlying matching coloured contours are the distributions of 1000 iterations accounting for the observational errors. The other filled symbols (circles, triangles) are the GCs members of the nearby progenitor families.}
    \label{fig:1000dist}
\end{figure}

\begin{figure}[h]
\centering
\includegraphics[width=\hsize]{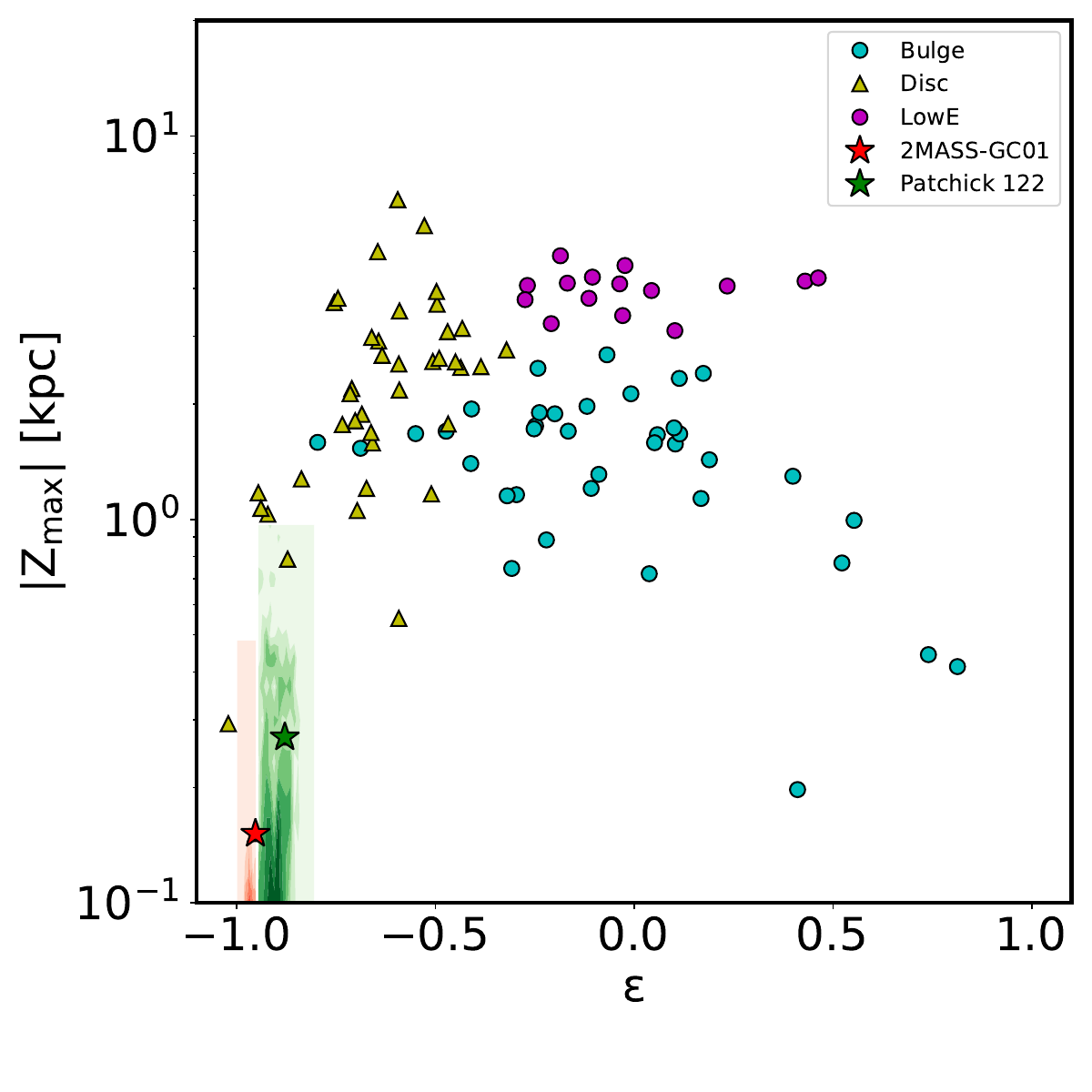}
\caption{$\varepsilon-Z_{max}$ space distribution of the GCs with the highest velocity vector uncertainties (filled stars). The underlying matching coloured contours are the distributions of 1000 iterations accounting for the observational errors. The other filled symbols (circles, triangles) are the GCs members of the nearby progenitor families.}
    \label{fig:1000vs}
\end{figure}

As mentioned in Sec.~\ref{data}, we tested that our results (the attribution of each GC to a specific progenitor) are robust even when taking into account observational errors. Here we show the worst cases out of the entire sample, even for them the classification doesn't become ambiguous when taking into account the uncertainties. In Fig.~\ref{fig:1000dist} we show the high energy GCs AM 1 (grey star), Eridanus (blue star), Palomar 3 (red star) and Palomar 4 (orange star), the GCs with the highest distance uncertainties, in the $L_z-L_{\perp}$ space. The underlying coloured contours in the same tonalities (greys for AM 1, blues for Eridanus, etc...) show the density distribution of 1000 integrations with different input phase space values generated by extracting from the error distributions of each GC. We also show the nearest progenitor families (Sag, HS, the retrogrades, and the other Elqui member). It is clear from the figure that, even for the cases with the largest scatter (AM 1, Palomar 4), accounting for the errors doesn't change the classification of these GCs. Similarly, in Fig.~\ref{fig:1000vs} we show 2MASS-GC01 (red star) and Patchick 122 (green star), the GCs with the highest uncertainties on their velocity vector, in $\varepsilon-Z_{max}$ space. The underlying matching coloured contours show the density distribution of 1000 iterations accounting for the observational errors. As before, we also show the nearest progenitor families (bulge, disc, and low-E) and it appears clear that, even accounting for the errors, the two GCs cannot be identified as anything but disc GCs.

\section{Sanity checks and tests of \textsc{OrbIT}}\label{appa}

\begin{figure}[h]
\centering
\includegraphics[width=\hsize]{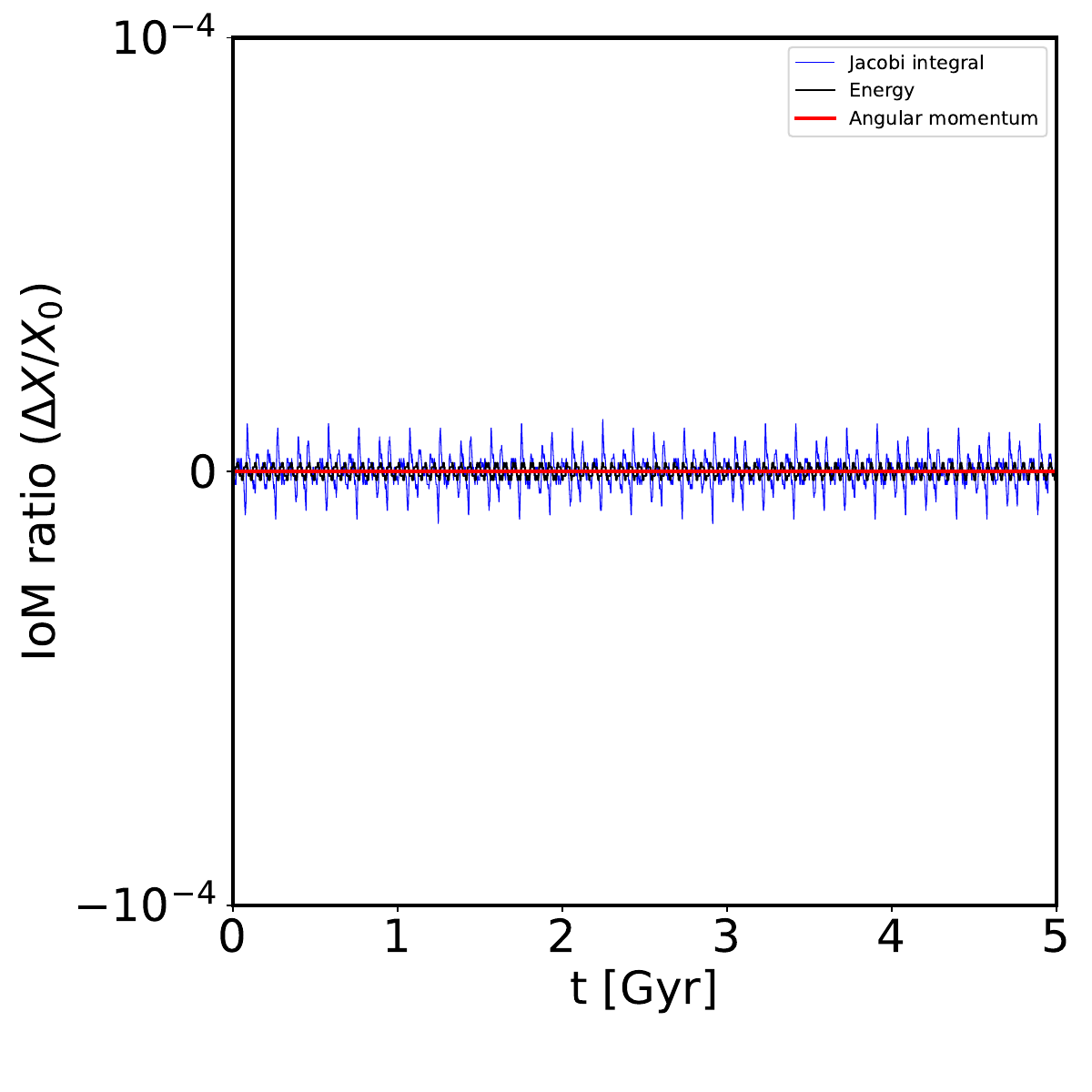}
\caption{Evolution of the IoMs over a typical orbital integration. In black and red, respectively, the total mechanical energy $E_{\rm tot}$ and the angular momentum $L_z$ for the case with a symmetric and static potential. In blue the Jacoby integral $E_J$ for the case with the full potential.}
    \label{fig:IoMs_ratios}
\end{figure}

To ensure that \textsc{OrbIT} works properly we performed an extensive array of tests, both in configurations with a symmetric and static potential (excluding the rotating bar and the central BH) and with the full time-evolving model. In the static and symmetric case, we checked the conservation of the 2 IoMs, the total mechanical energy $E_{\rm tot} = K + U$ and the angular momentum around the symmetry axis, $L_z$. With the full model, the only surviving IoM is the Jacoby energy \citep[or Jacobi integral][]{2008gady.book.....B}, $E_J=E_{\rm tot}-\Omega_b \cdot L$ (where $\Omega_b$ is the pattern speed of the rotation and $L$ is the angular momentum corresponding to the axis of rotation, in our case $L_z$).
Fig.~\ref{fig:IoMs_ratios} shows the ratios of the variation of the three IoMs during integrations of $5\cdot10^6$ steps for a total of $5 \ Gyr$. The ratios are computed as $\Delta X / X_0$ with the variation of each IoM being the value at every step minus the initial value, $\Delta X = X - X_0$ (with $X$ being one of $E_{\rm tot}, \ L_z, \ E_J$).

\FloatBarrier

\section{Data table}\label{appb}

Table ~\ref{tab2} is an excerpt of the results derived by \textsc{OrbIT}, the full table is only available in electronic form at the CDS..

\begin{table}
\centering
\resizebox{0.885\textwidth}{!}{\begin{minipage}{\textwidth}
\captionsetup{justification=centering}
  \caption{Recovered orbital parameters, IoMs, adiabatic invariants, and derived quantities used for this analysis for each GC studied. The full table is available at the CDS.}\label{tab2}
  \begin{tabular}{| c | c | c | c | c | c | c | c | c | c | c |}
 \hline
 Cluster & $R_{peri}$ & $R_{apo}$ & $ecc$ & $\lvert Z_{max} \rvert$ & $E_{tot}$ & $L_z$ & $L_{\perp}$ & $J_{\parallel}$ & $J_{\perp}$ & $\varepsilon$  \\
  & ${\rm [kpc]}$ & ${\rm [kpc]}$ &  & ${\rm [kpc]}$ & ${\rm [ km^2/s^2]}$ & ${\rm [kpc \ km/s]}$ & ${\rm [kpc \ km/s]}$ &  &  & \\
 \hline
 2MASS-GC01 & $2.54 \pm 0.90$ & $4.90 \pm 0.21$ & $0.34 \pm 0.15$ & 0.15 & -148604.9 & -924.6027 & 62.08 & -0.96 & -0.04 & -0.95 \\
 \hline
 2MASS-GC02 & $1.19 \pm 0.59$ & $2.79 \pm 0.28$ & $0.43 \pm 0.25$ & 1.17 & -195646.9 & -217.38 & 84.90 & -0.68 & -0.22 & -0.51 \\
 \hline
 AM 1 & $38.08 \pm 13.46$ & $167.96 \pm 59.38$ & $0.84 \pm 0.30$ & 110.59 & -32665.03 & 1273.80 & 8558.58 & 0.10 & 0.27 & 0.10 \\    
 \hline
 AM 4 & $26.45 \pm 2.04$ & $37.72 \pm 2.91$ & $0.19 \pm 0.02$ & 34.41 & -59019.18 & 1034.26 & 5417.18 & 0.18 & 0.77 & 0.18 \\
 \hline
 Arp 2 & $21.50 \pm 3.63$ & $95.34 \pm 16.12$ & $0.74 \pm 0.12$ & 78.61 & -42849.35 & -1611.98 & 5775.14 & -0.19 & 0.21 & -0.17 \\
 \hline
 BH 140 & $1.87 \pm 0.32$ & $11.68 \pm 1.00$ & $0.74 \pm 0.03$ & 2.65 & -118729.4 & -721.66 & 297.19 & -0.53 & -0.38 & -0.45 \\
 \hline
 BH 261 & $1.69 \pm 0.34$ & $2.64 \pm 0.23$ & $0.22 \pm 0.13$ & 1.05 & -183308.9 & -372.60 & 243.23 & -0.77 & 0.07 & -0.70 \\
 \hline
 Crater & $116.34 \pm 67.17$ & $220.74 \pm 127.45$ & $0.46 \pm 0.27$ & 128.76 & -25927.12 & 5541.70 & 14870.94 & 0.31 & 0.48 & 0.32 \\
 \hline
 Djor 1 & $1.31 \pm 0.28$ & $8.50 \pm 0.73$ & $0.74 \pm 0.04$ & 1.60 & -133058.6 & -544.76 & 173.68 & -0.53 & -0.42 & -0.44 \\
 \hline
 Djor 2 & $0.60 \pm 0.04$ & $0.78 \pm 0.02$ & $0.13 \pm 0.04$ & 0.44 & -242466.5 & 129.15 & 94.67 & 0.77 & 0.14 & 0.74 \\
  \hline
 \end{tabular}
\end{minipage}}
\end{table}

\end{appendix}
\end{document}